\input amstex
\magnification 1200
\TagsOnRight
\def\qed{\ifhmode\unskip\nobreak\fi\ifmmode\ifinner\else
 \hskip5pt\fi\fi\hbox{\hskip5pt\vrule width4pt
 height6pt depth1.5pt\hskip1pt}}
\NoBlackBoxes \baselineskip 18.7 pt
\parskip 6 pt
\def\stretch {\noalign{\medskip}}
\define \bC {\bold C}
\define \bCp {\bold C^+}
\define \bCm {\bold C^-}
\define \bCpb {\overline{\bold C^+}}
\define \bCmb {\overline{\bold C^-}}
\define \ds {\displaystyle}
\define \bR {\bold R}
\define \bm {\bmatrix}
\define \endbm {\endbmatrix}

\centerline {\bf INVERSE SCATTERING ON THE HALF LINE}
\centerline {\bf FOR THE MATRIX SCHR\"ODINGER EQUATION}

\vskip 5 pt
\centerline {Tuncay Aktosun}
\vskip -8 pt
\centerline {Department of Mathematics}
\vskip -8 pt
\centerline {University of Texas at Arlington}
\vskip -8 pt
\centerline {Arlington, TX 76019-0408, USA}
\vskip -8 pt
\centerline {aktosun\@uta.edu}

\centerline {Ricardo Weder\plainfootnote{$^\dagger$}
{Fellow Sistema Nacional de Investigadores}}
\vskip -8 pt
\centerline {Departamento de F\'\i sica Matem\'atica}
\vskip -8 pt
\centerline {Instituto de Investigaciones en
Matem\'aticas Aplicadas y en Sistemas}
\vskip -8 pt
\centerline {Universidad Nacional Aut\'onoma de M\'exico}
\vskip -8 pt
\centerline {Apartado Postal 20-126, IIMAS-UNAM, M\'exico DF 01000, M\'exico}
\vskip -8 pt \centerline {weder\@unam.mx}

\vskip 3 pt
\centerline {\it ``Dedicated to the 95th birthday of Prof. Vladimir A. Marchenko"}
\vskip 10 pt

\noindent {\bf Abstract}: The matrix Schr\"odinger equation is considered
on the half line with the general selfadjoint boundary condition at
the origin described by two boundary matrices satisfying
certain appropriate conditions. It is assumed that the matrix
potential is integrable, is selfadjoint, and has a finite
first moment. The corresponding scattering
data set is constructed, and such scattering data sets are characterized
by providing a set of necessary and sufficient conditions
assuring the existence and uniqueness of the one-to-one correspondence
between the scattering data set
and the input data set containing the potential and boundary matrices.
The work presented here provides a generalization of the classic
result by Agranovich and Marchenko from the Dirichlet boundary condition
to the general selfadjoint
boundary condition.




\newpage

\noindent {\bf 1. INTRODUCTION}
\vskip 3 pt

Our aim in this paper is to describe the direct and inverse scattering
problems for the half-line matrix Schr\"odinger operator with a selfadjoint boundary condition.
In the direct problem we are given an input data set $\bold D$
consisting of an $n\times n$ matrix-valued potential $V(x)$ and
a selfadjoint boundary condition at $x=0,$ and our goal is to determine the corresponding scattering data set $\bold S$ consisting of the scattering matrix $S(k)$ and
the bound-state data. In the inverse problem, we are given a scattering
data set $\bold S,$ and our goal is to determine the corresponding input data set
$\bold D.$ We would like to have a one-to-one correspondence between
an input data set $\bold D$ and a scattering data set $\bold S$
so that both the direct and inverse problems are well posed.
Thus, some restrictions are needed on $\bold D$ and $\bold S$
for a one-to-one correspondence.

Since the scattering and inverse scattering problems in the scalar case, i.e. when
$n=1,$ are well understood, it is desirable that the analysis in the matrix case reduces
to the scalar case when $n=1.$ However, as elaborated in
 Section~8, the current formulation of the
scattering and inverse scattering problems in the scalar case
presents a problem. As a consequence, it becomes impossible to have a one-to-one correspondence between an input data set $\bold D$ and a scattering data set $\bold S,$ unless the Dirichlet and non-Dirichlet boundary conditions are analyzed
separately and they are not mixed with each other. Although not ideal, this could perhaps be done in the scalar case because a given boundary condition in the scalar case is either Dirichlet or non-Dirichlet. On the other hand, in the matrix case with $n\ge 2,$
a given boundary condition may partly be Dirichlet and partly non-Dirichlet, and this may be as a result of constraints in a physical problem. It turns out that
the proper way to deal with
the issue is to modify the definition of the scattering matrix in such a way that
it is defined the same way regardless of the boundary condition, i.e. one should
avoid defining the scattering matrix in one way in the Dirichlet case and in another
way in the non-Dirichlet case.

There are four aspects related to the direct and inverse problems. These are the existence,
uniqueness, construction, and characterization. In the existence
aspect in the direct problem, given $\bold D$ in a specified class
we determine whether a corresponding $\bold S$ exists in some
specific class. The uniqueness
aspect is concerned with whether there exists a unique $\bold S$ corresponding
to a given $\bold D,$ or two or more distinct sets $\bold S$
may correspond to the same $\bold D.$
The construction deals with
the recovery of $\bold S$ from $\bold D.$
In the inverse
problem the existence problem deals with the existence of some
$\bold D$ corresponding to a given $\bold S$ belonging to a particular class.
The uniqueness deals with the question whether $\bold D$ corresponding to
a given $\bold S$ is
unique, and the construction consists of the recovery of
$\bold D$ from $\bold S.$
After the existence and uniqueness aspects in the direct and
inverse problems are settled, one then turns the attention to the
characterization problem, which consists of
the identification of the class to which $\bold D$ belongs
and the identification of the class to which $\bold S$ belongs so that
there is a one-to-one correspondence between $\bold D$ and $\bold S$
in the respective classes.
One also needs to ensure that the scattering data set $\bold S$ uniquely constructed from a given $\bold D$
in the direct problem in turn uniquely constructs the same $\bold D$
in the inverse problem.

A viable characterization in the literature for the matrix Schr\"odinger operator
on the half line can be found in the seminal work by Agranovich and
Marchenko [1]. However, the
analysis in [1] is restricted to the Dirichlet boundary condition,
and hence our study can be viewed as a generalization of
the characterization in [1]. A characterization for the case of the
general selfadjoint boundary condition is recently provided [9,10]
by the authors of this paper, and in the current paper we present a summary of
some of the results in [9,10]. For brevity, we do not include any proofs because such proofs
are already available in [9,10].

We present the existence, uniqueness, reconstruction, and
characterization
issues related to the relevant direct and inverse problems under the assumption
that $\bold D$ belongs to the {\it Faddeev class}
and $\bold S$ belongs to the {\it Marchenko class}.
The Faddeev class consists of input data sets
$\bold D$ as in (2.1), where the potential $V(x)$ and the
boundary matrices $A$ and $B$ are as specified in Definition~2.1.
The Marchenko class consists of scattering data sets $\bold S$
as in (3.12), where
the scattering matrix $S(k)$ and the bound-state data
$\{\kappa_j,M_j\}_{j=1}^N$ are as specified in Definition~4.1.

Let us mention the relevant references [14-16], where
the direct and inverse problems for (2.2)
are formally studied with the general selfadjoint boundary
condition, not as in (2.5)-(2.7) but in a form equivalent to (2.5)-(2.7). However,
the study in [14-16] lacks the large-$k$ analysis beyond the
 leading term and also lacks the small-$k$ analysis of the scattering data,
 which are both essential for the
analysis of the relevant inverse problem. Thus, our study can also
be considered as a complement to the work by Harmer [14-16].
In our paper, which is essentially a brief summary of [9,10],
 we rely on results from
previous work [1,4,5,8,22-14], in particular [1,4,8,22].

Our paper is organized as follows. In Section~2 we introduce
the matrix Schr\"odinger equation on the half line, describe the general selfadjoint
boundary condition in terms of two constant matrices $A$ and $B.$
We then describe the Faddeev class of input data sets $\bold D$ consisting of
the matrix potential $V(x)$ and the boundary matrices $A$ and $B.$
In Section~3 we describe the solution to the direct problem,
which uses an input data set $\bold D$ in the Faddeev class.
We outline the construction
of various quantities such as the Jost solution, the physical solution,
the regular solution, the Jost matrix, the scattering matrix, and the
bound-state data. In Section~4 we introduce
the Marchenko class of scattering data sets. We
present the solution to
the inverse problem by starting with a scattering data set
$\bold S$ in the Marchenko class, and we describe the construction
of the potential and the boundary matrices.
In Section~5 we provide a characterization of the
scattering data by showing that there is a one-to-one
correspondence between the Faddeev class of input data sets $\bold D$
and the Marchenko class of scattering data sets $\bold S.$
In Section~6 we provide an equivalent description of the Marchenko class,
and we provide an alternate characterization of the
scattering data with the help of Levinson's theorem.
In Section~7 we provide yet another description of the Marchenko class
based on an approach utilizing the so-called generalized Fourier
map. Finally, in Section~8 we contrast our definition of the
Jost matrix and the scattering matrix with those definitions
in the previous literature. We indicate the similarities and
differences occurring when the boundary condition used is
Dirichlet or non-Dirichlet. We
elaborate
on the resulting nonuniqueness issue if the scattering matrix is defined
differently when the Dirichlet boundary condition is used,
as commonly done in the previous literature.

\vskip 10 pt

\noindent {\bf 2. THE MATRIX SCHR\"ODINGER EQUATION}
\vskip 3 pt

In this section we introduce the
matrix Schr\"odinger equation (2.2), the matrix potential $V(x),$ and the
boundary matrices $A$ and $B$ used to describe the general selfadjoint boundary condition.
We also indicate that the boundary matrices $A$ and $B$ can be uniquely specified
modulo a postmultiplication by an invertible matrix. Our input data set $\bold D$
is defined as
$$\bold D:=\{V,A,B\}.\tag 2.1$$

Consider the matrix Schr\"odinger equation on the half line
$$-\psi''+V(x)\,\psi=k^2\psi,\qquad x\in\bR^+,\tag 2.2$$
where $\bR^+:=(0,+\infty),$ the prime denotes the derivative with respect to
the spatial coordinate $x,$
$k^2$ is the complex-valued spectral parameter,
the potential $V(x)$ is an $n\times n$ selfadjoint matrix-valued function of $x$ and
belongs to class $L_1^1(\bR^+),$ and $n$ is any positive integer.
We assume that the value of
$n$ is fixed and is known.
The selfadjointness of $V(x)$ is expressed as
$$V(x)=V(x)^\dagger,\qquad x\in\bR^+,\tag 2.3$$
where the dagger denotes the matrix adjoint (complex
conjugate and matrix transpose).
We equivalently say hermitian to describe a selfadjoint matrix.
We remark that, unless we are in the scalar case, i.e. unless $n=1,$
the potential is not necessarily real valued.
The condition $V\in L_1^1(\bR^+)$
means that
each entry of the matrix $V(x)$ is Lebesgue measurable on
$\bR^+$ and
$$\int_0^\infty dx\,(1+x)\,|V(x)|<+\infty,\tag 2.4$$
where $|V(x)|$ denotes the matrix operator norm.
Clearly, a matrix-valued function belongs
to $L^1_1(\bR^+)$
if and only if each entry of that matrix belongs
to $L^1_1(\bR^+).$

The wavefunction $\psi(k,x)$ appearing in (2.2)
may be either an $n\times n$ matrix-valued function
or it may be a column vector with $n$
components. We use $\bC$ for the complex plane,
$\bR$ for the real line $(-\infty,+\infty),$
$\bR^-$ for the left-half line $(-\infty,0),$
$\bCp$ for the open upper-half complex plane,
$\bCpb$ for $\bCp\cup\bR,$ $\bCm$ for the open lower-half complex plane, and
$\bCmb$ for $\bCm\cup\bR.$

We are interested in studying (2.2) with an $n\times n$
 selfadjoint matrix potential $V(x)$
in $L^1_1(\bR^+)$ under the general
selfadjoint boundary condition at $x=0.$
There are various equivalent formulations [4,8,14-18] of
the general
selfadjoint boundary condition at $x=0,$ and we find it
convenient
to state it [4,8] in terms of two
constant $n\times n$ matrices $A$ and $B$ as
$$-B^\dagger\psi(0)+A^\dagger\psi'(0)=0,\tag 2.5$$
where $A$ and $B$ satisfy
$$-B^\dagger A+A^\dagger B=0,\tag 2.6$$
$$A^\dagger A+B^\dagger B>0.\tag 2.7$$
The condition in (2.7) means that
the $n\times n$ matrix $(A^\dagger
A+B^\dagger B)$ is positive, which is also called positive definite.
One can easily verify that (2.5)-(2.7) remain invariant if
the boundary matrices $A$ and $B$
are replaced with $AT$ and $BT,$ respectively,
where $T$ is an arbitrary $n\times n$ invertible matrix.
We express this fact by saying that
the selfadjoint boundary condition (2.5) is uniquely determined
by the matrix pair $(A,B)$ modulo an invertible
matrix $T,$ and we equivalently state
 that (2.5) is equivalent to
the knowledge of $(A,B)$ modulo $T.$
We remark that the positivity condition (2.7) is equivalent to having the
rank of the $2n\times n$ matrix $\bm A\\ B\endbm$ equal to $n.$

In our analysis of the direct problem
related to (2.2) and (2.5), we
assume that our input data set $\bold D$ belongs to the Faddeev class
defined below.


\noindent {\bf Definition 2.1}
{\it
The input data set $\bold D$ given in
(2.1) is said to belong to the Faddeev class if
the potential $V(x)$ satisfies (2.3) and (2.4) and the boundary matrices
$A$ and $B$ appearing
in (2.5) satisfy (2.6) and (2.7).
In other words,
$\bold D$ belongs to the Faddeev class if the
$n\times n$ matrix-valued
potential $V(x)$ appearing in (2.2)
is hermitian and belongs to class
$L^1_1(\bR^+)$ and the constant $n\times n$
matrices $A$ and $B$ appearing in
(2.5) satisfy (2.6) and (2.7).}


It is possible to formulate the general selfadjoint boundary condition
by using a unique $n\times n$ constant matrix instead of using
the pair of matrices $A$ and $B$ appearing in (2.5)-(2.7).
For example, in [15] a unitary $n\times n$
matrix $U$ is used to describe the selfadjoint
boundary condition as
$$\ds\frac{i}{2}\left(U^\dagger -I\right)\psi(0)+\ds\frac{1}{2}
\left(U^\dagger +I\right)\psi'(0)=0,$$
where $I$ is the $n\times n$ identity matrix.
Without loss of any generality, one
could also use [4] a diagonal representation of the selfadjoint boundary condition
by choosing the matrices $A$ and $B$ as
$$A=\text{diag}\{-\sin \theta_1,\cdots,-\sin \theta_n\}
,\quad
B=\text{diag}\{\cos \theta_1,\cdots,\cos \theta_n\},\tag 2.8
$$
where the $\theta_j$ are some real constants in the
interval $(0,\pi].$ In fact, through the representation
(2.8), one can directly identify [5] the three integers
$n_D,$ $n_N,$ and $n_M,$ where $n_D$ is the number of
$\theta_j$-values equal to $\pi,$ $n_N$ is the number of
$\theta_j$-values equal to $\pi/2,$ and $n_M$
is the number of $\theta_j$-values in the union
$(0,\pi/2)\cup (\pi/2,\pi).$ One can informally call
$n_D$ the number of Dirichlet boundary conditions,
$n_N$ the number of Neumann boundary conditions, and
$n_M$ the number of mixed boundary conditions.

We find it more convenient to write the general selfadjoint
boundary condition in terms of the two
constant $n\times n$ matrices $A$ and $B,$ with the understanding that
$A$ and $B$ are unique up to a postmultiplication by an invertible
matrix $T$. For example, the so-called Kirchhoff boundary condition is
easier to recognize if expressed in terms of $A$ and $B,$ rather than
written in terms of a single unique $n\times n$ constant matrix.

\vskip 10 pt
\noindent {\bf 3. THE SOLUTION TO THE DIRECT PROBLEM}
\vskip 3 pt

In this section we summarize the solution to the
direct scattering problem associated with
(2.2) and (2.5) when the related input data set $\bold D$ given in
(2.1) belongs to the Faddeev class. In other words, we start with
an $n\times n$ hermitian potential $V(x)$ belonging to
$L^1_1(\bR^+)$ and a pair of constant boundary matrices $A$ and
$B$ satisfying (2.6) and (2.7), and we construct the relevant
quantities leading to the scattering data set $\bold S.$
The unique construction of the scattering data set $\bold S$ also
enables us to determine the basic properties of $\bold S.$
The steps of the construction are given below:

\item
{(a)}
When our input data set $\bold D$ belongs to the Faddeev class,
regardless of the boundary matrices
$A$ and $B,$ the
matrix Schr\"odinger equation (2.2) has an
$n\times n$ matrix-valued solution, usually called the Jost solution and denoted by
$f(k,x),$
satisfying the asymptotic condition
$$f(k,x)=e^{ikx} [I+o(1)],\qquad x\to+\infty.\tag 3.1$$
The solution $f(k,x)$ is uniquely determined by
the potential $V(x).$ For each fixed $x\in[0,+\infty),$
the Jost solution $f(k,x)$ has an extension from $k\in\bR$ to
$k\in\bCpb,$ and such an extension is continuous in $k\in\bCpb$ and
analytic in $k\in\bCp$ and
has the asymptotic behavior
$$e^{-ikx}f(k,x)=I +o(1),\qquad k\to\infty \text{ in } \bCpb.$$

\item
{(b)}
In terms of the boundary matrices $A$ and $B$ in
$\bold D$ and the Jost solution $f(k,x)$
obtained as in (a), we construct
the Jost matrix $J(k)$
as
$$J(k):=f(-k^\ast,0)^\dagger B-f'(-k^\ast,0)^\dagger A,\qquad k\in\bR,\tag 3.2$$
where the asterisk denotes complex conjugation.
We remark that $J(k)$ is an $n\times n$ matrix-valued function of $k.$
The redundant appearance of $k^\ast$
instead of $k$ in (3.2) when $k\in\bR$ is useful
in extending the Jost matrix analytically from
$k\in\bR$ to $k\in\bCpb.$
We recall that the boundary matrices $A$ and $B$ can be postmultiplied
by any invertible matrix $T$ without affecting (2.5)-(2.7)
and hence the definition given in (3.2)
yields the Jost matrix $J(k),$ which is unique up to a postmultiplication
by $T.$

\item
{(c)}
In terms of the Jost matrix $J(k),$
obtained from $\bold D$ as indicated in (3.2),
we construct the scattering matrix $S(k)$
as
$$S(k):=-J(-k)\,J(k)^{-1},\qquad k\in\bR.\tag 3.3$$
We remark that $S(k)$ is an $n\times n$ matrix-valued function of $k.$
Even though the Jost matrix in (3.2)
is uniquely determined up to a postmultiplication
by an invertible matrix $T,$ from (3.3) we see that
the scattering matrix $S(k)$ is uniquely determined
irrespective of $T.$

\item
{(d)}
In terms of the Jost solution $f(k,x)$
obtained in (a) and the scattering matrix $S(k)$
obtained in (c), we construct the so-called
physical solution to (2.2). The physical solution, denoted by
$\Psi(k,x),$ is constructed as
$$\Psi(k,x):=f(-k,x)+f(k,x)\,S(k),\qquad k\in\bR.\tag 3.4$$
We remark that $\Psi(k,x)$ is an $n\times n$ matrix-valued function of $k$ and $x.$
The physical solution, as the name implies, has the physical
interpretation of a scattering solution; namely, the initial
$n\times n$ matrix-valued
plane wave $e^{-ikx}I$
sent from $x=+\infty$ onto the potential yields the
$n\times n$ matrix-valued
scattered wave
$S(k)\,e^{ikx}$ at $x=+\infty$ with the amplitude $S(k).$ This interpretation is
seen by using (3.1) in (3.4), i.e. for each fixed $k\in\bR,$ we get
$$\Psi(k,x)=e^{-ikx}+S(k)\,e^{ikx}+o(1),\qquad x\to+\infty.$$
We also remark that each column of
the physical solution satisfies the boundary condition
(2.5), and hence the physical solution itself satisfies (2.5) and we have
$$-B^\dagger \Psi(k,0)+A^\dagger \Psi'(k,0)=0.\tag 3.5$$
Even though the boundary matrices $A$ and $B$ appearing
in (2.5)-(2.7) can be modified by a postmultiplication by
an invertible matrix $T,$
the definition given in
(3.4) uniquely determines the physical solution
irrespective of $T.$ In the definition of the physical
solution (3.4), one could multiply the right-hand side of
(3.4) by a scalar function of $k$ without affecting the
physical interpretation of a physical solution.
Nevertheless, we prefer to use (3.4) to define
the physical solution in a unique manner.

\item
{(e)}
Instead of constructing the physical solution
via (3.4), one can alternatively construct it in an equivalent way as follows.
When our input data set $\bold D$ belongs to the Faddeev class, there exists [4]
an $n\times n$ matrix-valued solution to (2.1), called the regular solution
and denoted by
$\varphi(k,x),$
satisfying the initial conditions
$$\varphi(k,0)=A,\quad \varphi'(k,0)=B.$$
The solution $\varphi(k,x)$ is uniquely determined
by the input data set $\bold D$ given in (2.1). We remark that
$\varphi(k,x)$ depends on the choice of $A$ and $B.$
The solution $\varphi(k,x)$ is known as
the regular solution because it is entire in
$k$ for each fixed $x\in\bR^+.$ In terms of the regular solution
$\varphi(k,x)$ and the Jost matrix $J(k)$ appearing in (3.2) we
can introduce the physical solution as
$$\Psi(k,x)=-2ik\,\varphi(k,x)\,J(k)^{-1}.\tag 3.6$$
One can show that the expressions given in (3.4) and (3.6)
are equivalent, and this can be shown by using the relationship given in
(3.5) of [4], i.e.
$$\varphi(k,x)=\ds\frac{1}{2ik}\,f(k,x)\,J(-k)-\ds\frac{1}{2ik}\,f(-k,x)\,J(k),\tag 3.7$$
where we recall that $f(k,x)$ is the Jost solution appearing in (3.1).

\item
{(f)}
When the input data set $\bold D$ belongs to
the Faddeev class,
the Jost matrix $J(k)$ constructed as in (3.2)
has an analytic extension from $k\in\bR$ to
$k\in\bCp$ and its determinant $\det[J(k)]$ is nonzero
in $\bCp$ except perhaps at a finite number of
$k$-values on the positive imaginary axis.
Let us use $N$ to denote the number of distinct zeros of
$\det[J(k)]$ in $\bCp$ without counting multiplicities of those zeros,
by realizing that the integer
$N$ may be zero for some input data sets $\bold D.$ Let us use $N$ distinct positive numbers
$\kappa_j$ so that the zeros of $\det[J(k)]$ occur at
$k=i\kappa_j$ and use $m_j$ to denote the multiplicity of
the zero of $\det[J(k)]$ at $k=i\kappa_j.$
Thus, the nonnegative integer $N,$ the set of distinct positive values
$\{\kappa_j\}_{j=1}^N,$
and the set of positive integers $\{m_j\}_{j=1}^N$ are all
uniquely determined by the input data set $\bold D.$
Each $m_j$ satisfies $1\le m_j\le n.$
It is appropriate to call $N$ the number of bound states without counting
the multiplicities. The nonnegative integer $\Cal N$ defined as
$$\Cal N:=\ds\sum_{j=1}^N m_j,\tag 3.8$$
can be referred to as the number of bound states including the multiplicities.

\item
{(g)}
Having determined the sets $\{\kappa_j\}_{j=1}^N$ and $\{m_j\}_{j=1}^N,$
let us use $\text{Ker}[J(i\kappa_j)^\dagger]$ to denote the kernel
of the $n\times n$ constant matrix $J(i\kappa_j)^\dagger.$ Next, we
construct the orthogonal projection
matrices $P_j$ onto $\text{Ker}[J(i\kappa_j)^\dagger]$ for $j=1,\dots,N.$
The $n\times n$ matrices $P_j$ are hermitian and idempotent, i.e.
$$P_j^\dagger=P_j,\quad P_j^2=P_j,\qquad j=1,\dots,N.$$
Furthermore, the rank of $P_j$ is equal to $m_j.$
We then construct the constant $n\times n$ matrices
$A_j,$ $B_j,$ and $M_j$ defined as
$$A_j:=\int_0^\infty dx\,f(i\kappa_j,x)^\dagger\, f(i\kappa_j,x),\qquad j=1,\dots,N,$$
$$B_j:=(I-P_j)+P_jA_j\,P_j,\qquad j=1,\dots,N,\tag 3.9$$
$$M_j:=B_j^{-1/2}P_j,\qquad j=1,\dots,N,\tag 3.10$$
where $f(k,x)$ is the Jost solution constructed in (a).
We remark that when $\bold D$ belongs to the Faddeev class, the matrices
$B_j$ given in (3.9) are hermitian and positive definite and hence the matrices
$B_j^{-1/2}$ are well defined as positive definite matrices. It is known that
each projection matrix $P_j$ has rank $m_j,$ and it follows from
(3.10) that each matrix $M_j$ is hermitian, nonnegative, and has rank
$m_j.$ The matrices $M_j$ are usually called the bound-state normalization matrices.

\item
{(h)}
When the input data set $\bold D$ belongs to the
Faddeev class, at each $k=i\kappa_j$ with $j=1,\dots,N,$
the Schr\"odinger equation (2.2) has $m_j$ linearly
independent column vector-valued solutions, where each of those column vector
solutions is square integrable in $x\in\bR^+.$
It is possible to rearrange those $m_j$ linearly
independent column-vector solutions to form
an $n\times n$ matrix
$\Psi_j(x),$ in such a way that
$\Psi_j(x)$ can be uniquely constructed as
$$\Psi_j(x):=f(i\kappa_j,x)\,M_j,\qquad j=1,\dots,N,\tag 3.11$$
where $M_j$ is the $n\times n$ normalization matrix
defined in (3.10). We can refer to $\Psi_j(x)$ as the
normalized bound-state matrix solution to
(2.1) at $k=i\kappa_j.$ We remark that each
$\Psi_j(x)$ satisfies the boundary condition (2.5)
and has rank equal to $m_j.$

Having constructed all the relevant quantities starting with the input data set
$\bold D,$ we now define the scattering data set $\bold S$ as
$$\bold S:=\{S,\{\kappa_j,M_j\}_{j=1}^N\},\tag 3.12$$
where $S$ denotes the scattering matrix
$S(k)$ for $k\in\bR$ constructed as
in (3.3), the $N$ distinct positive constants $\kappa_j$ are
as described in (f), and the $N$ hermitian, nonnegative, rank-$m_j$
matrices $M_j$ are as in (3.10).


\vskip 10 pt
\noindent {\bf 4. THE SOLUTION TO THE INVERSE PROBLEM}
\vskip 3 pt

In this section, given the scattering
data set $\bold S$ in (3.12), our goal is to construct
the input data set $\bold D$ given in (2.1),
with the understanding that the potential $V(x)$
is uniquely constructed and that
the boundary matrices $A$ and $B$ are uniquely constructed up to a postmultiplication
by an invertible matrix. The construction
is given when $\bold S$ belongs to the so-called Marchenko class.
We first present the construction and provide the definition
of the Marchenko class
at the end of the construction procedure.
Later in the section we show that the Marchenko class can also
be described in various equivalent ways.

We summarize the steps in the construction of $\bold D$ from $\bold S$ as follows:


\item
{(a)}
 From the large-$k$ asymptotics of the
scattering matrix $S(k),$ we determine the constant
$n\times n$ matrix $S_\infty$ via
$$S_\infty:=\lim_{k\to\pm\infty} S(k),\tag 4.1$$
and the constant
$n\times n$ matrix $G_1$ via
$$S(k)=S_\infty+\ds\frac{G_1}{ik}+o\left(\ds\frac{1}{k}\right),\qquad k\to\pm\infty.\tag 4.2$$

\item
{(b)}
Using $S(k)$ and $S_\infty,$
we uniquely construct
the $n\times n$ matrix $F_s(y)$ via
$$F_s(y):=\ds\frac{1}{2\pi} \int_{-\infty}^\infty dk\,[S(k)-S_\infty]\,e^{iky},\qquad y\in\bR.\tag 4.3$$

\item
{(c)}
Using $F_s(y)$ constructed as in
(4.3) and the bound-state data $\{\kappa_j,M_j\}_{j=1}^N$
appearing in $\bold S,$ we construct the $n\times n$ matrix
$F(y)$ via
$$F(y):=F_s(y)+\ds\sum_{j=1}^N M_j^2 \,e^{-\kappa_j y},\qquad y\in\bR^+.\tag 4.4$$
Note that we have $F_s(y)$ for $y\in\bR,$ but we
need $F(y)$ only for $y\in\bR^+.$

\item
{(d)}
We use the matrix $F(y)$ given in (4.4) as input to the
Marchenko integral equation
$$K(x,y)+F(x+y)+\int_x^\infty dz\,K(x,z)\,F(z+y)=0,\qquad 0\le x<y,\tag 4.5$$
and uniquely solve (4.5) and obtain $K(x,y)$
for $0\le x<y<+\infty.$ We remark that $K(x,y)$
is continuous in the region
$0\le x<y<+\infty.$ We note that $K(0,0),$
which is used to denote $K(0,0^+),$ is well
defined as a constant
$n\times n$ matrix.

\item
{(e)}
Having obtained $K(x,y)$ for $0\le x<y<+\infty$
uniquely from $\bold S$ as described in (d),
we construct the potential $V(x)$ via
$$V(x)=-2\,\ds\frac{d K(x,x)}{dx},\qquad x\in\bR^+.\tag 4.6$$
By $K(x,x)$ we mean $K(x,x^+).$ We remark that,
in general, $V(x)$ constructed as in (4.6) may exists only
a.e. and it may not be continuous in $x.$

\item
{(f)}
Having constructed the potential $V(x)$ from
the scattering data set $\bold S,$ we turn our attention to the
construction of the boundary matrices $A$ and $B$ appearing in
(2.1). We recall that we need to construct $A$ and $B$
uniquely, where the uniqueness is understood in the sense
of being up to a postmultiplication by an arbitrary invertible
$n\times n$ matrix $T.$ Such a construction is carried out as follows.
We use the already-constructed $n\times n$ constant matrices
$S_\infty,$ $G_1,$ and $K(0,0)$ as input in the linear, homogeneous algebraic system
$$\cases (I-S_\infty)\,A=0,\\ \stretch (I+S_\infty)\,B=\left[G_1-S_\infty\,K(0,0)-K(0,0)\,S_\infty\right]A,\endcases\tag 4.7$$
and determine $A$ and $B$ as the general solution to
(4.7).
Such a general solution
is equivalent to finding $A$ and $B$ satisfying
 (4.7) in such a way that the rank of the $2n\times n$ matrix $\bm A \\ B\endbm$ is
equal to $n.$

Even though the steps outlined above complete the construction of
the input data set $\bold D$ from the scattering data set $\bold S,$
we can construct various auxiliary quantities relevant
to the corresponding direct and inverse scattering problems as follows.

\item
{(g)}
Having constructed the solution $K(x,y)$ to the
Marchenko integral equation (4.5), we obtain the
Jost solution $f(k,x)$ via
$$f(k,x)=e^{ikx}I+\ds\int_x^\infty dy\, K(x,y)\,e^{iky}.\tag 4.8$$

\item
{(h)}
Having the Jost solution $f(k,x)$ and the scattering matrix
$S(k)$ at hand, we construct the physical solution $\Psi(k,x)$ as in
(3.4).

\item
{(i)}
Having the Jost solution $f(k,x)$ and the boundary matrices $A$ and $B$
at hand, we construct
Jost matrix $J(k)$ as in (3.2). Note that the constructed $A$ and $B$
are unique up to a postmultiplication by an arbitrary invertible matrix $T,$ and hence
the constructed Jost matrix $J(k)$ is also unique up to a postmultiplication by $T.$

\item
{(j)}
Having the Jost solution $f(k,x)$ and the Jost matrix $J(k)$ at hand,
we construct the regular solution $\varphi(k,x)$ as in (3.7). Since
 the constructed $A$ and $B$ as well as the constructed $J(k)$
are each unique up to a postmultiplication by an arbitrary invertible matrix $T,$
the constructed regular solution
 $\varphi(k,x)$ is also unique up to a postmultiplication by $T.$ For each particular choice of the pair $(A,B),$ we have a particular choice of the regular solution.

\item
{(k)}
Having the Jost solution $f(k,x)$ and the bound-state data $\{\kappa_j,M_j\}_{j=1}^N$
appearing in $\bold S,$ we construct the normalized bound-state
matrix solutions $\Psi_j(x)$ as in (3.11).


Next we define the Marchenko class of scattering data sets $\bold S.$
The importance of the Marchenko class is that there exists
[9,10] a one-to-one correspondence
between the Faddeev class of input data sets $\bold D$ and the Marchenko
class of scattering data sets $\bold S.$

\noindent {\bf Definition 4.1}
{\it Consider a scattering data set $\bold S$
as in (3.12), which consists of
an $n\times n$ scattering matrix $S(k)$ for $k\in\bR,$ a set of $N$ distinct
 positive constants $\kappa_j,$ and a set of
$N$ constant $n\times n$ hermitian and nonnegative matrices
$M_j$ with respective positive ranks $m_j,$ where $N$ is a nonnegative integer.
In case $N=0,$ it is understood that $\bold S$ consists only of
$S(k)$ for $k\in\bR.$
We say that $\bold S$ belongs to the Marchenko class if $\bold S$
satisfies the following four conditions, listed below as
$(\bold 1),$ $(\bold 2),$ $(\bold 3_a),$
$(\bold 4_a)$:}


\item {$(\bold 1)$} {\it The scattering matrix $S(k)$ satisfies}
$$S(-k)=S(k)^\dagger=S(k)^{-1},\qquad k\in\bR,\tag 4.9$$
{\it and there exist constant $n\times n$ matrices
$S_\infty$ and $G_1$ in such a way that (4.2) holds. Furthermore,
the $n\times n$ matrix quantity $F_s(y)$ defined in (4.3) is
bounded in $y\in\bR$ and integrable in $y\in\bR^+.$}

\item {$(\bold 2)$}  {\it For the matrix $F_s(y)$ defined in
(4.3), the derivative $F_s'(y)$ exists a.e. for $y\in\bR^+$ and it satisfies}
$$\int_0^\infty dy\,(1+y)\,|F'_s(y)|<+\infty,\tag 4.10$$
{\it where we recall that the norm in the integrand of (4.10) is the
operator norm of a matrix.}

\item {$(\bold 3_a)$}  {\it The physical solution $\Psi(k,x)$ satisfies
the boundary condition (2.5), i.e. it satisfies (3.5).
We clarify this
property as follows: The scattering matrix appearing in $\bold S$ yields a
particular $n\times n$
matrix-valued solution $\Psi(k,x)$
to (2.2) known as the physical solution given in (3.4)
and also yields a pair of matrices $A$ and $B$ (modulo an invertible
matrix) satisfying (2.6) and (2.7).
Our statement $(\bold 3_a)$ is equivalent
to saying that (2.5) is satisfied if we use in (2.5) the quantities
$\Psi(k,x),$ $A,$ and $B$ constructed from $\bold S.$}

\item {$(\bold 4_a)$}  {\it The Marchenko equation (4.5) at $x=0$ given by
$$K(0,y)+F(y)+\int_0^\infty dz\,K(0,z)\,F(z+y)=0,\qquad y\in\bR^+,$$
has a unique solution $K(0,y)$ in
$L^1(\bR^+).$
Here, $F(y)$ is the $n\times n$
matrix related
to $F_s(y)$ as in (4.4).}


Let us mention a slight drawback in the definition of the
Marchenko class given in Definition~4.1. The property
$(\bold 3_a)$ cannot be checked from the scattering data set
$\bold S$ directly because it requires the construction of the corresponding
boundary matrices $A$ and $B$ as well as the physical solution
$\Psi(k,x).$ It is already known [9,10] that one can replace
$(\bold 3_a)$ by an equivalent pair of
 conditions, listed as $(\bold{III}_a)$ and $(\bold V_c),$
as indicated in the next theorem.


\noindent {\bf Theorem 4.2} {\it
Consider a scattering data set $\bold S$
as in (3.12), which consists of
an $n\times n$ scattering matrix $S(k)$ for $k\in\bR,$ a set of $N$ distinct
 positive constants $\kappa_j,$ and a set of
$N$ constant $n\times n$ hermitian and nonnegative matrices
$M_j$ with respective positive ranks $m_j,$ where $N$ is a nonnegative integer.
In case $N=0,$ it is understood that $\bold S$ consists only of
$S(k)$ for $k\in\bR$ and that $\Cal N$ appearing in (3.8) is zero.
The scattering data set $\bold S$ belongs to the Marchenko class if and only if
$\bold S$
satisfies the five conditions, three of which are listed as
$(\bold 1),$ $(\bold 2),$ and
$(\bold 4_a)$ in Definition~4.1, and
the two additional conditions $(\bold{III}_a)$ and $(\bold V_c)$
are given by:}


\item {$(\bold {III}_a)$} {\it For the matrix-valued function $F_s(y)$ given in (4.3),
the derivative
$F_s'(y)$ for $y\in\bR^-$
can be written as a sum of two matrix-valued functions, one of which
is integrable and the other is square integrable in $y\in\bR^-.$
Furthermore, the only solution
$X(y),$ which is a row vector with $n$ square-integrable
components in $y\in\bR^-,$ to the linear
homogeneous integral equation}
$$-X(y)+\int_{-\infty}^0 dz\,X(z)\,F_s(z+y)=0,\qquad
y\in\bR^-,$$
{\it is the trivial solution $X(y)\equiv 0.$}

\item {$(\bold {V}_c)$}  {\it The linear homogeneous integral equation}
$$X(y)+\int_0^\infty dz\,X(z)\,F_s(z+y)=0,\qquad y\in\bR^+,\tag 4.11$$
{\it has precisely $\Cal N$ linearly independent row-vector
solutions for some nonnegative integer
$\Cal N,$ with $n$ components which are integrable in $y\in\bR^+.$
Here $F_s(y)$ is the matrix defined in (4.3) and $\Cal N$ is
the nonnegative integer readily constructed from $\bold S$ as in (3.8).
If $\Cal N=0,$ it is understood that the only solution in $L^1(\bR^+)$ to
(4.11) is the trivial solution $X(y)\equiv 0.$}



We remark that Theorem~4.2 is a special case of
Theorem~4.5, but we still prefer
to state it as a separate result. This is because
Theorem~4.2 is closely related to the characterization
result stated by Agranovich and Marchenko
in the Dirichlet case on pp. 4--5 of
their manuscript [1].

The next theorem shows that in the description of the Marchenko class
specified in Definition~4.1, we can replace the condition~$(\bold 3_a)$
with another equivalent condition.


\noindent {\bf Theorem 4.3} {\it
Consider a scattering data set $\bold S$
as in (3.12), which consists of
an $n\times n$ scattering matrix $S(k)$ for $k\in\bR,$ a set of $N$ distinct
 positive constants $\kappa_j,$ and a set of
$N$ constant $n\times n$ hermitian and nonnegative matrices
$M_j$ with respective positive ranks $m_j,$ where $N$ is a nonnegative integer.
In case $N=0,$ it is understood that $\bold S$ consists only of
$S(k)$ for $k\in\bR.$
The scattering data set $\bold S$ belongs to the Marchenko class if and only if
$\bold S$
satisfies the four conditions, three of which are listed as
$(\bold 1),$ $(\bold 2),$ and
$(\bold 4_a)$ in Definition~4.1 and
one additional condition $(\bold 3_b)$ replacing
$(\bold 3_a)$, which is given by}


\item {$(\bold 3_b)$}
{\it
 The Jost matrix $J(k)$ satisfies
$$J(-k)+S(k)\,J(k)=0,\qquad k\in\bR.\tag 4.12$$
{\it We clarify this property as follows:
The scattering matrix $S(k)$ given in $\bold S$
yields a Jost matrix
$J(k)$ constructed as in (3.2), unique up to a postmultiplication by an invertible matrix.
Using the scattering matrix $S(k)$ given in
$\bold S$ and the Jost matrix constructed from $S(k),$ we find that (4.12)}
is satisfied.}


Let us use $\hat L^1(\bCp)$ to denote the Banach space of all complex-valued functions
$\xi(k)$
that are analytic in $k\in\bCp$ in such a way that
there exists a corresponding function $\eta(x)$ belonging to
$L^1(\bR^+)$ satisfying
$$\xi(k)=\int_0^\infty dx\,\eta(x)\, e^{ikx}.$$
We remark that
if $\xi(k)$ belongs to $\hat L^1(\bCp),$ then $\xi(k)$
is continuous in $k\in\bR$ and it satisfies $\xi(k)=o(1)$
as $k\to\infty$ in $\bCpb.$ If $\xi(k)$ is vector valued or matrix valued
instead of being scalar valued, then it belongs to $\hat L^1(\bCp)$ if and only if each entry of $\xi(k)$ belongs to $\hat L^1(\bCp).$

We remark that the result of
Theorem~4.3 is included
in the next theorem presented.
However, we have stated
Theorem~4.3 separately in order
to emphasize
the importance of
$(\bold 3_b)$ of Theorem~4.3
and its connection to $(\bold 3_a)$ of Definition~4.1.


\noindent {\bf Theorem 4.4} {\it
Consider a scattering data set $\bold S$
as in (3.12), which consists of
an $n\times n$ scattering matrix $S(k)$ for $k\in\bR,$ a set of $N$ distinct
 positive constants $\kappa_j,$ and a set of
$N$ constant $n\times n$ hermitian and nonnegative matrices
$M_j$ with respective positive ranks $m_j,$ where $N$ is a nonnegative integer.
In case $N=0,$ it is understood that $\bold S$ consists only of
$S(k)$ for $k\in\bR.$
The scattering data set $\bold S$ belongs to the Marchenko class if and only if
$\bold S$
satisfies the four conditions $(\bold 1),$ $(\bold 2),$
$(\bold 3),$ and $(\bold 4),$
where
$(\bold 3)$ can be either one of $(\bold 3_a)$ and $(\bold 3_b);$
and $(\bold 4)$ can be any one of $(\bold 4_a),$ $(\bold 4_b),$ $(\bold 4_c),$
$(\bold 4_d),$ and $(\bold 4_e).$
Note that $(\bold 1)$, $(\bold 2)$, $(\bold 3_a),$ $(\bold 4_a)$
are listed in Definition~4.1;
$(\bold 3_b)$ is listed in Theorem~4.3;
and the remaining conditions $(\bold 4_b)$,
$(\bold 4_c)$, $(\bold 4_d)$, $(\bold 4_e)$
are listed below:}
%


\item {$(\bold 4_b)$}  {\it The only solution in $L^1(\bR^+)$ to the
homogeneous Marchenko integral equation at $x=0$ given by}
$$K(0,y)+\int_0^\infty dz\,K(0,z)\,F(z+y)=0, \qquad y\in\bR^+,\tag 4.13$$
{\it is the trivial solution $K(0,y)\equiv 0.$
Note that (4.13) is the homogeneous version at $x=0$ of
the Marchenko
equation given by (4.5).
We remark that $F(y)$ appearing in (4.13) is the quantity defined in (4.4).}

\item {$(\bold 4_c)$} {\it The only integrable solution
$X(y),$ which is a row vector with $n$
integrable components in $y\in\bR^+,$ to the linear homogeneous integral equation}
$$X(y)+\int_0^\infty dz\,X(z)\,F(z+y)=0,\qquad y\in\bR^+,\tag 4.14$$
{\it is the trivial solution $X(y)\equiv 0.$
Again, we recall that $F(y)$ is the quantity defined in (4.4).}

\item {$(\bold 4_d)$}
{\it The only solution $\hat X(k)$ to the system}
$$\cases \hat X(i\kappa_j)\,M_j=0,\qquad j=1,\dots,N,\\
\stretch
\hat X(-k)+\hat X(k)\,S(k)=0,\qquad k\in\bR,\endcases\tag 4.15$$
{\it where $\hat X(k)$ is a row vector
with $n$ components belonging
to the class $\hat L^1(\bCp),$ is the trivial
solution $\hat X(k)\equiv 0.$}

\item {$(\bold 4_e)$} {\it
The only solution $h(k)$ to the system}
$$\cases M_j\, h(i\kappa_j)=0,\qquad j=1,\dots,N,\\
\stretch
h(-k)+S(k)\,h(k)=0,\qquad k\in\bR,\endcases\tag 4.16$$
{\it where $h(k)$ is a column vector
with $n$ components belonging
to the class $\hat L^1(\bCp),$ is the trivial
solution $h(k)\equiv 0.$}



We use $\bold H^2(\bC^\pm)$ to denote the Hardy space of all complex-valued functions $\xi(k)$
that are analytic in $k\in\bC^\pm$
with a finite norm defined as
$$||\xi||_{\bold H^2(\bC^\pm)}:=\sup_{\rho >0}\left[\int_{-\infty}^\infty d\alpha\,|\xi(\alpha\pm i\rho)|^2\right]^{1/2}.$$
Thus, $\xi(k)$ is square integrable along all lines in $\bC^\pm$ that are
parallel to the real axis.
The value of $\xi(k)$ for $k\in\bR$ is defined to be
the non-tangential limit of $\xi(k\pm i\rho)$ as $\rho \to 0^+.$
Such a non-tangential limit exists a.e. in $k\in\bR$ and belongs to
$L^2(\bR).$
It is known that $\xi(k)$ belongs to $\bold H^2(\bCp)$ if and only if
there exists a corresponding function $\eta(x)$ belonging to $L^2(\bR^+)$ in such a way that
$$\xi(k)=\int_0^\infty dx\, \eta(x)\, e^{ikx}.$$
Similarly, $\xi(k)$ belongs to $\bold H^2(\bCm)$ if and only if
there exists a corresponding function $\eta(x)$ belonging to $L^2(\bR^-)$
in such a way that
$$\xi(k)=\int_{-\infty}^0 dx\, \eta(x)\, e^{ikx}.$$
If $\xi(k)$ is vector valued or matrix valued
instead of being scalar valued, then it belongs to
$\bold H^2(\bC^\pm)$ if and only if each entry of $\xi(k)$ belongs to
$\bold H^2(\bC^\pm).$

The next theorem shows that in the equivalent description of the Marchenko class
specified in Theorem~4.2, we can replace the condition~$(\bold {III}_a)$
with one of two other equivalent conditions and we can also
replace the condition~$(\bold V_c)$ with any one of
various other equivalent conditions.


\noindent {\bf Theorem 4.5} {\it
Consider a scattering data set $\bold S$
as in (3.12), which consists of
an $n\times n$ scattering matrix $S(k)$ for $k\in\bR,$ a set of $N$ distinct
 positive constants $\kappa_j,$ and a set of
$N$ constant $n\times n$ hermitian and nonnegative matrices
$M_j$ with respective positive ranks $m_j,$ where $N$ is a nonnegative integer.
In case $N=0,$ it is understood that $\bold S$ consists only of
$S(k)$ for $k\in\bR$ and that $\Cal N$ appearing in (3.8) is zero.
The scattering data set $\bold S$ belongs to the Marchenko class if and only if
$\bold S$
satisfies the five conditions
$(\bold 1),$ $(\bold 2),$ $(\bold{III}),$
$(\bold 4),$ and $(\bold V),$ where
$(\bold{III})$ represents any one of
the three conditions $(\bold{III}_a),$ $(\bold{III}_b),$ $(\bold{III}_c);$
$(\bold 4)$ represents any one of the five conditions
$(\bold 4_a),$ $(\bold 4_b),$ $(\bold 4_c),$ $(\bold 4_d),$
$(\bold 4_e);$
and $(\bold V)$
represents any one of the eight conditions $(\bold V_a),$ $(\bold V_b),$
$(\bold V_c),$ $(\bold V_d),$ $(\bold V_e),$ $(\bold V_f),$ $(\bold V_g),$ $(\bold V_h).$
We remark that $(\bold 1),$ $(\bold 2),$ and
$(\bold 4_a)$ are listed in Definition~4.1;
 $(\bold{III}_a)$ and $(\bold V_c)$
 are listed in Theorem~4.2;
 $(\bold 4_b),$ $(\bold 4_c),$ $(\bold 4_d),$
$(\bold 4_e)$ are listed in Theorem~4.4;
and the remaining conditions
are listed below:}


\item {$(\bold {III}_b)$} {\it For the matrix-valued function $F_s(y)$ given in (4.3),
the derivative
$F_s'(y)$ for $y\in\bR^-$
can be written as a sum of two matrix-valued functions, one of which
is integrable and the other is square integrable in $y\in\bR^-.$
 Furthermore, the only solution $\hat X(k)$ to the homogeneous Riemann-Hilbert problem
$$-\hat X(-k)+\hat X(k)\,S(k)=0,\qquad k\in\bR,$$
where $\hat X(k)$ is a row vector
with $n$ components belonging
to the class $\bold H^2(\bCm),$ is the trivial
solution $\hat X(k)\equiv 0.$}

\item {$(\bold {III}_c)$} For the matrix-valued function $F_s(y)$ given in (4.3),
the derivative
$F_s'(y)$ for $y\in\bR^-$
can be written as a sum of two matrix-valued functions, one of which
is integrable and the other is square integrable in $y\in\bR^-.$
Furthermore, the only solution $h(k)$ to the homogeneous Riemann-Hilbert problem
$$-h(-k)+S(k)\, h(k)=0,\qquad k\in\bR,\tag 4.17$$
where $h(k)$ is a column vector
with $n$ components belonging
to the class $\bold H^2(\bCm),$ is the trivial
solution $h(k)\equiv 0.$

\item {$(\bold {V}_a)$} {\it Each of the $N$ normalized bound-state matrix solutions $\Psi_j(x)$
constructed as in (3.11) satisfies the boundary condition (2.5), i.e.}
$$-B^\dagger \Psi_j(0)+A^\dagger \Psi'_j(0)=0,\qquad j=1,\dots,N.\tag 4.18$$
{\it We clarify this statement as follows. The scattering matrix $S(k)$ and the bound-state data $\{\kappa_j,M_j\}_{j=1}^N$
given in $\bold S$ yield $n\times n$ matrices $\Psi_j(x)$ as in
(3.11),
where each $\Psi_j(x)$ is a solution to
(2.2) at $k=i\kappa_j.$
As stated
in $(\bold 3_a)$ of Definition~4.1, the scattering matrix given in $\bold S$ yields a pair of matrices $A$ and $B$ (modulo an invertible
matrix) satisfying (2.6) and (2.7). The statement $(\bold {V}_a)$ is equivalent
to saying that (2.5) is satisfied if we use in (2.5) the quantities
$\Psi_j(x),$ $A,$ and $B$ constructed from the quantities appearing in $\bold S.$
If $N=0,$ then the condition (4.18) is absent.}

\item {$(\bold {V}_b)$} {\it The normalization matrices
$M_j$ appearing in $\bold S$ satisfy}
$$J(i\kappa_j)^\dagger M_j=0,\qquad j=1,\dots,N.\tag 4.19$$
{\it We clarify this condition as follows. As indicated in $(\bold 3_b)$
of Theorem~4.3,
the scattering matrix $S(k)$ given in $\bold S$
yields a Jost matrix
$J(k).$ Using in (4.19) the matrix $M_j$ given in
$\bold S$ and the Jost matrix constructed from $S(k),$ at each
$\kappa_j$-value listed in $\bold S$ the matrix
equation (4.19) holds. If $N=0,$ then the condition (4.19) is absent.}

\item {$(\bold {V}_d)$}  {\it The homogeneous Riemann-Hilbert problem given by}
$$\hat X(-k)+\hat X(k)\,S(k)=0,\qquad k\in\bR,\tag 4.20$$
{\it has precisely $\Cal N$ linearly independent
row-vector solutions with $n$ components in
$\hat L^1(\bCp).$
Here, $\Cal N$ is the nonnegative integer given in (3.8).
If $\Cal N=0,$ it is understood that the only solution in $\hat L^1(\bCp)$ to
(4.20) is the trivial solution $\hat X(k)\equiv 0.$}

\item {$(\bold {V}_e)$} {\it The homogeneous Riemann-Hilbert problem given by}
$$h(-k)+S(k)\,h(k)=0,\qquad k\in\bR,\tag 4.21$$
{\it has precisely $\Cal N$ linearly independent
column-vector solutions with $n$ components in $\hat L^1(\bCp),$ where
 $\Cal N$ is the nonnegative integer given in (3.8).
If $\Cal N=0,$ it is understood that the only solution in $\hat L^1(\bCp)$ to
(4.21) is the trivial solution $h(k)\equiv 0.$}

\item {$(\bold {V}_f)$} {\it The integral equation (4.11)
has precisely $\Cal N$ linearly independent
 row-vector solutions
$X(y)$ with $n$ components in $L^2(\bR^+),$ where $\Cal N$ is the nonnegative integer
given in (3.8). If $\Cal N=0,$ it is understood that the only solution in $L^2(\bR^+)$ to
(4.11) is the trivial solution $X(y)\equiv 0.$
We remark that the matrix $F_s(y)$
appearing in the kernel of (4.11)
is defined in (4.3).}

\item {$(\bold {V}_g)$} {\it The homogeneous Riemann-Hilbert problem given
in (4.20)
has precisely $\Cal N$ linearly independent
row-vector solutions $\hat X(k)$ with
$n$ components in $\bold H^2(\bCp),$
Here, $\Cal N$ is the nonnegative integer given in (3.8).
If $\Cal N=0,$ it is understood that the only solution in $\bold H^2(\bCp)$ to
(4.20) is the trivial solution $\hat X(k)\equiv 0.$}

\item {$(\bold {V}_h)$} {\it The homogeneous Riemann-Hilbert problem given
in (4.21)
has precisely $\Cal N$ linearly independent
row-vector solutions with $n$ components in $\bold H^2(\bCp).$
Here, $\Cal N$ is the nonnegative integer given in (3.8).
If $\Cal N=0,$ it is understood that the only solution in $\bold H^2(\bCp)$ to
(4.21) is the trivial solution $h(k)\equiv 0.$}



\vskip 10 pt
\noindent {\bf 5. THE CHARACTERIZATION OF THE SCATTERING DATA}
\vskip 3 pt

In this section we consider the characterization of
the scattering data.
In the next theorem we present one of our main characterization results.
It shows that
the four conditions given in Definition~4.1 for the Marchenko class
form a characterization of the scattering data sets $\bold S$
so that
there exists a one-to-one
correspondence between a scattering data set
in the Marchenko class and an input data set $\bold D$
in the Faddeev class specified in Definition~2.1.
 From Section~4 we know that the Marchenko class can be described in various
 equivalent ways, and hence it is possible to present the characterization
 in various different ways.


\noindent {\bf Theorem 5.1} {\it
Consider a scattering data set $\bold S$
as in (3.12), which consists of
an $n\times n$ scattering matrix $S(k)$ for $k\in\bR,$ a set of $N$ distinct
 positive constants $\kappa_j,$ and a set of
$N$ constant $n\times n$ hermitian and nonnegative matrices
$M_j$ with respective positive ranks $m_j,$ where $N$ is a nonnegative integer.
In case $N=0,$ it is understood that $\bold S$ consists only of
$S(k)$ for $k\in\bR$ and that $\Cal N$ appearing in (3.8) is zero.
Consider also an input data set
$\bold D$ as in (2.1) consisting of an $n\times n$ matrix potential
$V(x)$ satisfying (2.3) and (2.4) and a pair of constant
$n\times n$ matrices $A$ and $B$ satisfying (2.6) and (2.7).
Then, we have the following:}


\item
{(a)} {\it
For each input data set $\bold D$ in the Faddeev class
specified in Definition~2.1, there exists and uniquely
exists a scattering data set $\bold S$ in the Marchenko class
specified in Definition~4.1.}

\item
{(b)} {\it
Conversely,
for each $\bold S$ in the Marchenko class, there exists
and uniquely
exists an input data set $\bold D$ in the Faddeev class,
where the boundary matrices $A$ and $B$ are uniquely determined up to a postmultiplication
by an invertible $n\times n$ matrix $T.$}

\item
{(c)} {\it
Let $\tilde {\bold S}$ be the scattering data set corresponding
to $\bold D$ given in the previous step (b), where $\bold D$ is constructed from
the scattering data set $\bold S.$ Then, we have
$\tilde {\bold S}=\bold S,$ i.e.
the scattering data set constructed from $\bold D$ is equal to
the scattering data set used to construct $\bold D.$}

\item
{(d)} {\it
The characterization outlined in the steps (a)-(c) given above
can equivalently be stated as follows.
A set $\bold S$ as in (3.12) is the scattering data set
corresponding to an input data set $\bold D$ in the Faddeev class
if and only if
$\bold S$ satisfies $(\bold 1)$, $(\bold 2)$, $(\bold 3_a)$, and $(\bold 4_a)$
stated in Definition~4.1.}

\item
{(e)} {\it
The characterization outlined in the steps (a)-(c) given above
 can equivalently be stated as follows.
A set $\bold S$ as in (3.12) is the scattering data set
corresponding to an input data set $\bold D$ in the Faddeev class
if and only if
$\bold S$ satisfies $(\bold 1)$, $(\bold 2)$, $(\bold 4_a)$ of
Definition~4.1
and $(\bold{III}_a)$ and $(\bold V_c)$
of Theorem~4.2.}

\item
{(f)} {\it
The characterization outlined in the steps (a)-(c) given above
 can equivalently be stated as follows.
A set $\bold S$ as in (3.12) is the scattering data set
corresponding to an input data set $\bold D$ in the Faddeev class
if and only if
$\bold S$ satisfies $(\bold 1)$, $(\bold 2)$, $(\bold 3),$ and $(\bold 4),$
where $(\bold 3)$ can be either one of $(\bold 3_a)$ and $(\bold 3_b);$
and $(\bold 4)$ can be any one of $(\bold 4_a),$ $(\bold 4_b),$ $(\bold 4_c),$
$(\bold 4_d),$ and $(\bold 4_e).$
We recall that $(\bold 1)$, $(\bold 2)$, $(\bold 3_a),$ $(\bold 4_a)$
are listed in Definition~4.1;
$(\bold 3_b)$ is listed in Theorem~4.3;
and $(\bold 4_b),$ $(\bold 4_c),$
$(\bold 4_d),$ $(\bold 4_e)$
are listed in Theorem~4.4.}

\item
{(g)} {\it
The characterization outlined in the steps (a)-(c) given above
 can equivalently be stated as follows.
A set $\bold S$ as in (3.12) is the scattering data set
corresponding to an input data set $\bold D$ in the Faddeev class
if and only if
$\bold S$ satisfies $(\bold 1)$, $(\bold 2)$, $(\bold {III}),$ $(\bold 4),$
and $(\bold V),$
where $(\bold {III})$ can be any one of
$(\bold {III}_a),$ $(\bold {III}_b),$ $(\bold {III}_c);$
$(\bold 4)$ can be any one of $(\bold 4_a),$ $(\bold 4_b),$ $(\bold 4_c),$
$(\bold 4_d),$ $(\bold 4_e);$
and $(\bold V)$ can be any one of $(\bold V_a),$ $(\bold V_b),$ $(\bold V_c),$
$(\bold V_d),$ $(\bold V_e),$ $(\bold V_f),$ $(\bold V_g),$ $(\bold V_h).$
We recall that $(\bold 1)$, $(\bold 2),$ $(\bold 4_a)$
are listed in Definition~4.1;
$(\bold {III}_a)$ and $(\bold V_c)$
are listed in Theorem~4.2;
$(\bold 4_b),$ $(\bold 4_c),$
$(\bold 4_d),$ and $(\bold 4_e)$
are listed in Theorem~4.4;
and $(\bold {III}_b),$ $(\bold {III}_c),$
$(\bold V_a),$ $(\bold V_b),$
$(\bold V_d),$ $(\bold V_e),$ $(\bold V_f),$ $(\bold V_g),$ and $(\bold V_h)$
are listed in Theorem~4.5.}


We have the following remarks on the results presented in
Theorem~5.1. The characterization result
stated in Theorem~5.1(e)
follows from Theorem~4.2.
The result stated in Theorem~5.1(f) is
a consequence of
Theorem~4.4.
The result in Theorem~5.1(e) is
a particular case of the
result in Theorem~5.1(g),
but we prefer to state it separately because it
resembles the characterization result stated by
Agranovich and Marchenko [1] in the Dirichlet case.
Finally we remark that
Theorem~5.1(g)
is a direct consequence of
Theorem~4.5.

\vskip 10 pt
\noindent {\bf 6. AN ALTERNATE CHARACTERIZATION OF THE SCATTERING DATA}
\vskip 3 pt

It is possible to present an alternate characterization of the scattering data
using Levinson's theorem. This characterization again establishes a one-to-one correspondence between the Faddeev class of input data sets $\bold D$ and
the Marchenko class of scattering data sets $\bold S.$ Hence,
such an alternate characterization can also be viewed as
an alternate description of the Marchenko class of scattering data sets
with the help of Levinson's theorem.

In general, the bound-state data $\{\kappa_j,M_j\}_{j=1}^N$ appearing in
the scattering data set $\bold S$ of (3.12) and the scattering matrix $S(k)$
are independent, and they need to be specified separately. On the other hand,
the determinant of the scattering matrix contains the information of
 the number of bound states including the multiplicities, which is the
nonnegative integer $\Cal N$ appearing in (3.8). The change in the argument of the determinant of
$S(k)$ as $k$ changes from $k=0^+$ to $k=+\infty$ in the $k$-interval $(0,+\infty)$
is related to the
number of bound states including multiplicities. This general fact
is usually known as Levinson's theorem.

When the input data set $\bold D$ belongs to the Faddeev class, we have [8] Levinson's
theorem stated in the following.


\noindent {\bf Theorem 6.1} {\it
Consider the matrix
Schr\"odinger equation (2.2) with the selfadjoint boundary condition
(2.5). Assume that the corresponding input data set
$\bold D$ given in (2.1) belongs to the Faddeev class. Let $\bold S$ appearing
in (3.12)
be the scattering data set corresponding
to $\bold D.$ Then, the number $\Cal N$
of bound states including the multiplicities
appearing in (3.8) is related to
the argument of the determinant of
the scattering matrix $S(k)$ as}
$$\text{arg}\left[\det [S(0^+)]\right]-\text{arg}\left[\det[ S(+\infty)]\right]=
\pi\left(2\Cal N+\mu+n_D-n
\right),\tag 6.1$$
{\it where $\mu$ is the (algebraic and geometric) multiplicity of
the eigenvalue $+1$ of the zero-energy
scattering matrix $S(0),$
$n$ is the positive integer appearing in the matrix size $n\times n$
of the scattering matrix $S(k),$
and
$n_D$ is the number of Dirichlet boundary conditions
in the diagonal representation (2.8) of the boundary matrices
$A$ and $B.$ We remark that
$n_D$ is the same as the nonzero integer
which is equal to the multiplicity of
the eigenvalue $-1$ of the
constant $n\times n$ matrix $S_\infty$ appearing in
(4.1).}


In some cases, by using Levinson's theorem we may be able to
quickly determine if a given scattering data set $\bold S$
does not belong to the Marchenko class. Using the
scattering matrix $S(k),$ we readily know the positive
integer $n$ appearing in the matrix size $n\times n$ of
the matrix $S(k).$ The zero-energy scattering matrix
$S(0$ has eigenvalues equal to either $-1$ or
$+1.$ Thus, we can identify $\mu$ as the multiplicity of the
eigenvalue $+1$ of $S(0).$ From the large-$k$ limit of
$S(k)$ given in (4.1) we can easily construct the constant matrix $S_\infty$
and we already know that $S_\infty$ has eigenvalues equal to either $-1$ or
$+1.$ Thus, we can identify $n_D$ as the multiplicity of
the eigenvalue $-1$ of $S_\infty.$ Then, from the scattering
matrix $S(k)$ we can evaluate the change in the argument of
$\det[S(k)]$ given on the left-hand side of (6.1).
We can then use (6.1) to determine the value
of $\Cal N$ predicted by Levinson's theorem. If that value of $\Cal N$ evaluated
 from (6.1) does not turn out to be a nonnegative integer, we know that
the corresponding $\bold S$ does not belong to the Marchenko class.

The next theorem shows that we can obtain an equivalent description of the Marchenko class
specified in Definition~4.1, by replacing $(\bold 3_a)$
by a set of two conditions one of which is related to Levinson's theorem, and at the same time
by replacing $(\bold 4_a)$ by any one of three other conditions.


\noindent {\bf Theorem 6.2} {\it
Consider a scattering data set $\bold S$
as in (3.12), which consists of
an $n\times n$ scattering matrix $S(k)$ for $k\in\bR,$ a set of $N$ distinct
 positive constants $\kappa_j,$ and a set of
$N$ constant $n\times n$ hermitian and nonnegative matrices
$M_j$ with respective positive ranks $m_j,$ where $N$ is a nonnegative integer.
In case $N=0,$ it is understood that $\bold S$ consists only of
$S(k)$ for $k\in\bR$ and that $\Cal N$ appearing in (3.8) is zero.
The scattering data set $\bold S$ belongs to the Marchenko class if and only if
$\bold S$
satisfies the five conditions, two of which are listed as
$(\bold 1)$ and $(\bold 2)$ in Definition~4.1,
the third and the fourth are the respective conditions
listed as $(\bold L)$ and $(\overset{\circ}\to {\bold 5})$ below, and the fifth
is any one of the three conditions listed as $(\overset{\circ}\to {\bold 4}_c), (\overset{\circ}\to {\bold 4}_d),$ $(\overset{\circ}\to {\bold 4}_e)$ below:}


\item {$(\bold L)$} {\it The scattering matrix $S(k)$
appearing in $\bold S$ is continuous for $ k \in \bold R$ and (6.1) of Levinson's theorem is satisfied with $\mu$, $n_D$, and $ \Cal N$
coming from $\bold S.$ Here,
$\mu$ is the (algebraic and geometric) multiplicity of the eigenvalue $+1$
  of the zero-energy scattering matrix $S(0)$, $n_D$ is the (algebraic and geometric) multiplicity of the eigenvalue $-1$ of  the hermitian matrix
   $S_\infty$ appearing in (4.1), and
   $ \Cal N$ is the nonnegative integer in (3.8) which is
   equal to the sum of the ranks $m_j$ of the matrices $M_j$
   appearing in $\bold S.$}

\item {$(\overset{\circ}\to {\bold 4}_c)$} {\it The only square-integrable solution $X(y),$ which is a row vector with $n$ square-integrable components in $ y \in \bold R^+$, to the linear homogeneous integral  equation}
$$X(y)+ \int_0^\infty\, dz \, X(z)\, F(z+y)=0,\qquad y\in\bR^+,\tag 6.2$$
{\it is the trivial solution $ X(y)\equiv 0$. Here, $F(y)$ is the quantity defined in (4.4).}

    \item {$(\overset{\circ}\to {\bold 4}_d)$}
{\it The only solution $\hat X(k)$ to the system}
$$\cases
\hat X(i\kappa_j)\,M_j=0,\qquad j=1,\dots,N,\\
\stretch
\hat X(-k)+\hat X(k)\,S(k)=0,\qquad k\in\bR,
\endcases
\tag 6.3$$
{\it where $\hat X(k)$ is a row vector
with $n$ components belonging
to the Hardy space $\bold H^2(\bCp),$ is the trivial
solution $\hat X(k)\equiv 0.$}

\item {$(\overset{\circ}\to {\bold 4}_e)$} {\it
The only solution $h(k)$ to the system}
$$
\cases
 M_j\, h(i\kappa_j)=0,\qquad j=1,\dots,N,\\
 \stretch
h(-k)+S(k)\,h(k)=0,\qquad k\in \bR,
\endcases
\tag 6.4$$
{\it where $h(k)$ is a column vector
with $n$ components belonging
to the Hardy space $\bold H^2(\bCp),$  is the trivial
solution $h(k)\equiv 0.$}

\item {$(\overset{\circ}\to {\bold 5})$} {\it
For the matrix-valued function $F_s(y)$ given in (4.3),
the derivative $F_s'(y)$ for $y\in\bR^-$ can be written as a sum 
of two matrix-valued functions, one of which is
integrable and the other is square integrable in $y\in\bR^-.$}



We remark that the conditions
$(\overset{\circ}\to {\bold 4}_c),$ $(\overset{\circ}\to {\bold 4}_d),$ $(\overset{\circ}\to {\bold 4}_e)$ listed
in Theorem~6.2 are somehow similar to
the conditions $(\bold 4_c),$ $(\bold 4_d),$ $(\bold 4e)$
of Theorem~4.4. However, there are also some differences; for example,
$X(y)$ appearing in (4.14) belongs to
$L^1(\bR^+)$ whereas $X(y)$ appearing in (6.2) belongs to $L^2(\bR^+),$
$\hat X(k)$ of (4.15) belongs to
$\hat L^1(\bCp)$ whereas $\hat X(k)$ of (6.3) belongs to
$\bold H^2(\bCp),$
and $h(k)$ of (4.16) belongs to
$\hat L^1(\bCp)$ whereas $h(k)$ of (6.4) belongs to
$\bold H^2(\bCp).$

Let us also remark that the condition
$(\overset{\circ}\to{\bold 5})$ in Theorem~6.2 is the same as the first sentence
given in $(\overset{\circ}\to{\bold {III}_a})$ of Theorem~4.2.
We note that Theorem~6.2 is the generalization
of a characterization result by
Agranovich and Marchenko presented in
Theorem~2 on p. 281 of [1],
which utilizes Levinson's theorem in the purely Dirichlet case.
That characterization result by
Agranovich and Marchenko is valid only in the case of
the Dirichlet boundary condition and does not include the condition
stated in $(\overset{\circ}\to{\bold 5})$ in Theorem~6.2.
In the special case of
the purely Dirichlet boundary condition,
it turns out that $(\overset{\circ}\to{\bold 5})$ in Theorem~6.2
is not needed. This has something to do with the fact that
in the purely Dirichlet case the Marchenko integral equation (4.5)
alone plays a key role in the solution to the inverse problem whereas
in the non-Dirichlet case not only the Marchenko integral equation
but also the derivative Marchenko integral equation plays a key role
in the solution to the inverse problem, in particular in the
satisfaction of the selfadjoint
boundary condition given in (3.5). The derivative Marchenko integral
equation is obtained by taking the $x$-derivative
of (4.5), and hence the quantity $F_s'(y)$ appears
in the nonhomogeneous term of the derivative Marchenko integral
equation. That presence of $F_s'(y)$ somehow results in
the condition stated in $(\overset{\circ}\to{\bold 5})$ in Theorem~6.2.
The presence of $(\overset{\circ}\to{\bold 5})$ in Theorem~6.2
also has something to do with the fact that
the boundary condition stated in (3.5) must hold for all
$k\in\bR.$ By taking the Fourier transform of
both sides of (3.5), we end up with the
requirement that
the Fourier transform of the left-hand side of
(3.5) must identically vanish. For this, one needs the necessity of
the satisfaction of $(\overset{\circ}\to{\bold 5})$ in Theorem~6.2,
unless $A=0$ in (3.5). Since the case $A=0$ is the same as
having the purely Dirichlet boundary condition, $(\overset{\circ}\to{\bold 5})$ in Theorem~6.2 is relevant only in the non-Dirichlet case.
For the mathematical elaboration on $(\overset{\circ}\to{\bold 5})$
we refer the reader to [10].

The presence of $(\overset{\circ}\to{\bold 5})$ in Theorem~6.2 is an indication of one of several reasons  why the characterization of scattering data sets with the general selfadjoint boundary condition is more involved than the characterization with the Dirichlet boundary condition.

We conclude that the result presented in Theorem~6.2, compared
to Theorem~5.1, constitutes an alternate characterization
of the scattering data sets $\bold S.$ Recall that Theorem~5.1 characterizes
the scattering data sets that are in a one-to-one correspondence
with the input data sets $\bold D$ in the Faddeev class. With the
help of Theorem~6.2 we have the following alternate
characterization of the scattering data sets.


\noindent {\bf Theorem 6.3} {\it
Consider a scattering data set $\bold S$
as in (3.12), which consists of
an $n\times n$ scattering matrix $S(k)$ for $k\in\bR,$ a set of $N$ distinct
 positive constants $\kappa_j,$ and a set of
$N$ constant $n\times n$ hermitian and nonnegative matrices
$M_j$ with respective positive ranks $m_j,$ where $N$ is a nonnegative integer.
In case $N=0,$ it is understood that $\bold S$ consists only of
$S(k)$ for $k\in\bR.$
Consider also an input data set
$\bold D$ as in (2.1) consisting of an $n\times n$ matrix potential
$V(x)$ satisfying (2.3) and (2.4) and a pair of constant
$n\times n$ matrices $A$ and $B$ satisfying (2.6) and (2.7),
where it is understood that the boundary matrices $A$ and $B$
are unique up to a postmultiplication by an invertible $n\times n$ matrix
$T.$
Then, we have the following
characterization of the scattering data sets.
A set $\bold S$ as in (3.12) is the scattering data set
corresponding to an input data set $\bold D$ in the Faddeev class
if and only if
$\bold S$ satisfies $(\bold 1)$ and $(\bold 2)$
of Definition~4.1, both $(\bold L)$ and $(\overset{\circ}\to{\bold 5})$
of Theorem~6.2, and
any one of the three conditions listed as $(\overset{\circ}\to {\bold 4}_c),$ $(\overset{\circ}\to {\bold 4}_d),$ $(\overset{\circ}\to {\bold 4}_e)$ in Theorem~6.2.}


\vskip 10 pt
\noindent {\bf 7. ANOTHER CHARACTERIZATION OF THE SCATTERING DATA}
\vskip 3 pt

In this section we provide yet another
 description of
the Marchenko class of scattering data sets $\bold S$
so that there exists a one-to-one correspondence
between an
input data set $\bold D$ in
the Faddeev class and a scattering data set $\bold S$
in the Marchenko class. Such a description allows us to have
yet another characterization of the scattering data sets
$\bold S$ in a one-to-one correspondence with the input data sets $\bold D$
in the Faddeev class.

The characterization given in this section resulting from a new description of the Marchenko class has some similarities and differences
compared to the first characterization presented in Theorem~5.1 and
the alternate characterization presented in Theorem~6.3.
Related to this
new characterization, the construction of the potential
in the solution to the inverse problem is the same as in the
previous characterizations; namely, one constructs the potential by solving the
Marchenko equation. Hence, the conditions $(\bold 1)$, $(\bold 2)$, $(\bold 4_a)$
in the first characterization, the conditions
$(\bold 1)$, $(\bold 2)$, $(\overset{\circ}\to {\bold 4}_c)$ in the alternate characterization,
and the conditions $(\bold I),$
$(\bold 2),$
$(\bold 4_c)$ in this new
characterization are essentially used to construct
the potential. This new characterization differs from
the two earlier ones in regard to the
satisfaction of the boundary condition by the physical solution
$\Psi(k,x)$ and by the normalized bound-state matrix solutions $\Psi_j(x).$
It is based on the alternate solution to the inverse problem by using the
generalized Fourier map [22].
This new characterization
uses six conditions,
indicated as $(\bold I),$
$(\bold 2),$ $(\bold A),$ $(\bold 4_c),$ either
of $(\bold V_e)$ or $(\bold V_h),$ and $(\bold {VI}).$
Recall that $(\bold 2)$ is described in Definition~4.1,
$(\bold 4_c)$ is described in Theorem~4.4, and $(\bold V_e)$ and $(\bold V_h)$
are described in Theorem~4.5. In the following definition we describe
the conditions $(\bold I),$ $(\bold A),$ and $(\bold {VI}).$


\noindent {\bf Definition 7.1}
{\it The properties $(\bold I)$, $(\bold A),$ and $(\bold {VI})$ for the scattering data set $\bold S$ in (3.12) are defined as follows:}


\item {$(\bold I)$} {\it The scattering matrix $S(k)$ satisfies
(4.9), the quantity $S_\infty$ appearing in
(4.1) exists, the quantity $S(k)-S_\infty$ is square integrable
in $k\in\bR,$ and the quantity $F_s(y)$ defined in
(4.3) is bounded in $y\in\bR$ and integrable in $y\in\bR^+.$}

\item {$(\bold A)$} {\it Consider
the nonhomogeneous
Riemann-Hilbert problem given by}
$$h(k)+S(-k)\, h(-k)=g(k),\qquad k\in\bR,\tag 7.1$$
{\it where
the nonhomogeneous
term $g(k)$
belongs to a dense
subset $\overset{\circ}\to \Upsilon$
of the vector space $\Upsilon$ of column vectors with $n$ square-integrable
components and satisfying $g(-k)=S(k)\,g(k)$ for $k\in\bR.$
Then, for each such given
$g(k),$ the equation (7.1) has a solution $h(k)$
as a column vector
with $n$ components belonging
to the Hardy space $\bold H^2(\bCp).$}

\item {$(\bold {VI})$}  {\it The scattering matrix $S(k)$ is continuous in $k\in\bR.$}


We remark that $(\bold I)$ of Definition~7.1 is weaker than $(\bold 1)$ of Definition~4.1. The quantity
$G_1$ and hence (4.2) appearing in $(\bold 1)$ are used to construct the boundary matrices
$A$ and $B$. In order to construct the potential $V(x)$ only, it is enough to use
the weaker condition $(\bold I).$
The condition $(\bold A)$  of Definition~7.1
somehow resembles $(\bold{III}_c)$ of Theorem~4.5, but there are also
some major differences.
In $(\bold{III}_c)$ a solution is sought to
the homogeneous Riemann-Hilbert problem (4.17)
as a column vector with $n$ components where each of those
components belongs to $\bold H^2(\bCm),$
and the only solution is expected to be the trivial solution
$h(k)\equiv 0.$
On the other hand, in
$(\bold A)$ one solves a nonhomogeneous
Riemann-Hilbert problem and the
solution is sought as a column vector where each of the
$n$ components belongs the Hardy space
$\bold H^2(\bCp).$ The solution
$h(k)$ to (7.1)
is in general nontrivial because
the nonhomogeneous term $g(k)$ there is in general
nontrivial, and the existence of a
solution to (7.1) is more relevant than its uniqueness.
The condition $(\bold {VI})$  of Definition~7.1,
which is the continuity of the scattering matrix $S(k),$
is mainly needed to prove that
the physical solution $\Psi(k,x)$ satisfies the boundary condition (2.5).

Let us first describe the solution to the inverse scattering problem
related to this new characterization and then present
the characterization itself. As already indicated,
the part of the solution to the inverse problem involving the construction of the
potential is practically the same as the solution outlined in Section~4.
However, the part of the solution related to the
boundary condition is different than the procedure outlined
in Section~4.
We summarize below the construction of $\bold D$ from $\bold S$ in this
new method,
where the existence and uniqueness are implicit at each step:


\item
{(a)}
 From the large-$k$ asymptotics of the
scattering matrix $S(k),$ with the help of
(4.1), we determine the $n\times n$ constant
matrix $S_\infty.$ Contrary to the method
of Section~4, we do not deal with the determination of
the constant
$n\times n$ matrix $G_1$ appearing in
(4.2).
It follows from (4.9) that the matrix
$S_\infty$ is hermitian when $\bold S$ satisfies
the condition $(\bold I)$ described in Definition~7.1.

\item
{(b)}
In terms of the quantities in $\bold S,$ we uniquely construct
the $n\times n$ matrix $F_s(y)$ by using (4.3) and
the $n\times n$ matrix $F(y)$ by using (4.4). This step is the same as
steps (b) and (c) of the summary of the method outlined in Section~4.

\item
{(c)}
If the condition $(\bold 4_c)$ of Theorem~4.4
is also satisfied,
then one uses the matrix $F(y)$ as input to the
Marchenko integral equation (4.5).
If $F(y)$ is integrable in $y\in(x,+\infty)$ for each $x\ge 0,$ then
for each fixed $x\ge 0$ there exists a solution $K(x,y)$ integrable
in $y\in(x,+\infty)$ to (4.5)
and such a solution is unique.
The solution $K(x,y)$ can be constructed by iterating (4.5).
We remark
that this step is the same as step (d) of the summary of the method outlined in Section~4.
Even though $K(x,y)$ is constructed only for $0\le x<y,$ one can extend
$K(x,y)$ to $y\in\bR^+$ by letting $K(x,y)=0$ for $0\le y<x.$

\item
{(d)}
Having obtained $K(x,y)$ uniquely from $\bold S,$ one
constructs the potential $V(x)$ via (4.6)
and also constructs the Jost solution
$f(k,x)$ via (4.8). Then,
by using $(\bold I),$ $(\bold 2)$, and $(\bold 4_c),$ one proves that the constructed $V(x)$ satisfies
(2.3) and (2.4) and that
the constructed $f(k,x)$ satisfies (2.2) used
with the constructed potential $V(x).$

\item
{(e)}
Having constructed
the Jost solution $f(k,x),$ one then
constructs the physical solution $\Psi(k,x)$ via
(3.4)and the normalized
bound-state matrix solutions
$\Psi_j(x)$ via (3.11). One then proves that the
constructed matrix $\Psi(k,x)$ satisfies (2.2)
and that the constructed $\Psi_j(x)$
satisfies (2.2) at $k=i\kappa_j,$
with the understanding that the constructed potential
$V(x)$ is used in (2.2).

\item
{(f)}
Having constructed the potential
$V(x),$ one forms a matrix-valued
differential operator denoted by ${\Cal L}_{\text{min}},$
which acts as $(-D_x^2 I+V)$ with $D_x:=d/dx,$
with a domain that is a dense subset
of $L^2(\bR^+).$ More precisely, the domain of ${\Cal L}_{\text{min}}$
consists of the column vectors with $n$ components
each of which is a function of $x$ belonging to
a dense subset of $L^2(\bR^+).$
The constructed
operator ${\Cal L}_{\text{min}}$ is symmetric, i.e.
it satisfies ${\Cal L}_{\text{min}}\subset {\Cal L}_{\text{min}}^\dagger,$
but is not selfadjoint, i.e. it
does not satisfy ${\Cal L}_{\text{min}}={\Cal L}_{\text{min}}^\dagger.$
The operator inclusion ${\Cal L}_{\text{min}}\subset {\Cal L}_{\text{min}}^\dagger$
indicates that the domain of the operator
${\Cal L}_{\text{min}}$ is a subset of
the domain of the operator ${\Cal L}_{\text{min}}^\dagger$
and these two operators have the same value on the
domain of ${\Cal L}_{\text{min}}.$

\item
{(g)}
One then constructs a selfadjoint realization of
${\Cal L}_{\text{min}},$ namely an operator ${\Cal L}$ in such a way that
${\Cal L}_{\text{min}}\subset {\Cal L}$ and ${\Cal L}={\Cal L}^\dagger.$
The constructed operator
${\Cal L}$ is a restriction of ${\Cal L}_{\text{min}}^\dagger,$ i.e.
we have ${\Cal L}\subset {\Cal L}_{\text{min}}^\dagger$
but not ${\Cal L}={\Cal L}_{\text{min}}^\dagger.$

\item
{(h)}
The construction of the operator ${\Cal L}$ is achieved [9,10,22]
by using the so-called generalized Fourier map $\bold F$ and its adjoint
$\bold F^\dagger.$ The generalized Fourier map
$\bold F$ corresponds to a generalization of the Fourier transform
between the space of square-integrable functions of $x$ and
the space of square-integrable functions of $k.$

\item
{(i)}
Once the selfadjoint operator ${\Cal L}$ is constructed,
it follows [9,10] that the domain of ${\Cal L}$
is a maximal isotropic subspace, which is sometimes also called a Lagrange plane.
Once we know that the domain of ${\Cal L}$ is a maximal isotropic subspace,
then it follows [9,10] that the functions in the
domain of ${\Cal L}$ must satisfy the boundary condition (2.5)
for some boundary matrices $A$ and $B$ satisfying
(2.6) and (2.7), where $A$ and $B$ are uniquely determined up to
a postmultiplication by an invertible matrix $T.$

\item
{(j)}
Finally, one proves that the constructed physical solution
$\Psi(k,x)$ and the constructed normalized
bound-state matrix solutions $\Psi_j(x)$ satisfy
the boundary condition (2.5) with the boundary matrices $A$ and
$B$ specified in the previous step; however, such a proof is
different in nature than the proofs for the previous characterizations. For the constructed matrices
$\Psi_j(x),$ it is immediate that they satisfy the
boundary condition because they belong to the domain of ${\Cal L}.$
Thus, it remains to prove that the constructed
$\Psi(k,x)$ satisfies the boundary condition.
We note that the matrix $\Psi(k,x)$ does not belong to
the domain of ${\Cal L}$ because its entries do not
belong to $L^2(\bR^+).$ On the other hand,
$\Psi(k,x)$ is locally square integrable in $x\in[0,+\infty),$ i.e.
it is square integrable in every compact subset of $[0,+\infty).$
Hence, it is possible to use a simple limiting argument to
prove that $\Psi(k,x)$ satisfies the boundary condition (2.5),
and the condition $(\bold {VI})$ is utilized
in the aforementioned limiting argument.

\item
{(k)}
As in the previous characterization given in Theorem~5.1(c), we still need to prove that
the input data set
$\bold D$ of (2.1) constructed from the scattering data set
$\bold S$ of (3.12) yields $\bold S.$
The proof of this step is the same as in the proof of Theorem~5.1(c).


Based on the procedure outlined above, we next present another
description of the Marchenko class of scattering data sets $\bold S.$
Recall that $(\bold I),$ $(\bold A),$ and
$(\bold {VI})$ are described in Definition~7.1, $(\bold 2)$
is described in Definition~4.1, $(\bold 4_c)$ is described in Theorem~4.4,
and $(\bold V_e)$ and $(\bold V_h)$ are described in Theorem~4.5.


\noindent {\bf Theorem 7.2} {\it
Consider a scattering data set $\bold S$
as in (3.12), which consists of
an $n\times n$ scattering matrix $S(k)$ for $k\in\bR,$ a set of $N$ distinct
 positive constants $\kappa_j,$ and a set of
$N$ constant $n\times n$ hermitian and nonnegative matrices
$M_j$ with respective positive ranks $m_j,$ where $N$ is a nonnegative integer.
In case $N=0,$ it is understood that $\bold S$ consists only of
$S(k)$ for $k\in\bR$ and that $\Cal N$ appearing in (3.8) is zero.
The set $\bold S$
is the scattering data set
corresponding to a unique input data set $\bold D$ as in (4.2)
in the Faddeev class
specified in Definition~2.1
if and only if $\bold S$ satisfies the six conditions consisting of
$(\bold I),$ $(\bold 2),$ $(\bold A),$ $(\bold 4_c),$ either one of
$(\bold V_e)$ and $(\bold V_h),$ and
$(\bold {VI}).$ We recall that the uniqueness of the input data set
$\bold D$ is understood in the sense that the boundary matrices
$A$ and $B$ in (4.2) are unique up to a postmultiplication by an arbitrary
invertible $n\times n$ matrix $T.$}


\vskip 10 pt
\noindent {\bf 8. SOME ELABORATIONS}
\vskip 3 pt

In this section we make a comparison with the definitions of
the Jost matrix and the scattering matrix in the scalar case
appearing in the literature. We also
elaborate on the nonuniqueness issue arising
if the scattering matrix is defined differently when the Dirichlet boundary condition is used. The reader is referred to Section~4 of [6] and Example~6.3 of
[6] for further elaborations on the nonuniqueness issue.

In the scalar case, i.e. when $n=1$, from (2.8) we see that we can choose
$$A=-\sin\theta, \quad B=\cos\theta,\qquad \theta\in(0,\pi],\tag 8.1$$
where $\theta$ represents the boundary parameter.
We can write the boundary condition (2.5) in the equivalent form
$$-A^\dagger \,\psi'(0)+B^\dagger \,\psi(0)=0.\tag 8.2$$
Using (8.1) in (8.2) we see that our boundary condition (2.5) in the scalar case
is equivalent to
$$(\sin\theta)\,\psi'(0)+(\cos\theta)\,\psi(0)=0,
\qquad \theta\in(0,\pi].\tag 8.3$$
We remark that the boundary condition (8.3) agrees with the boundary
condition used in the literature [7,19,20] in the scalar case. Since
$\theta=\pi$ corresponds to the Dirichlet boundary condition, we
can write (8.3) in the equivalent form
$$\cases \psi(0)=0,\qquad \text{Dirichlet case},\\
\stretch
\psi'(0)+(\cot \theta)\,\psi(0)=0,\qquad \text{non-Dirichlet case},\endcases\tag 8.4$$
where $\theta\in(0,\pi)$ in the non-Dirichlet case.
The boundary condition (8.4) is also identical [7,19,20]
to that used in the literature in the scalar case.
As stated below (2.7), the boundary matrices $A$ and $B$ in (2.5)
can be postmultiplied by any invertible matrix $T$
without affecting (2.5)-(2.7). Hence,
the constants $A$ and $B$ appearing in (8.1)
can be multiplied by any nonzero constant. In any case, the boundary
condition (2.5) we use is the same as the boundary condition
used in the literature [7,19,20] in the scalar case.

Using (8.1) we see that the Jost matrix defined in (3.2) yields
$$J(k)=f(-k^\ast,0)^\dagger \,(\cos\theta)+f'(-k^\ast,0)^\dagger
\,(\sin\theta),\qquad k\in\bR,\tag 8.5$$
where we recall that $\theta=\pi$ in the Dirichlet case
and $\theta\in(0,\pi)$ in the non-Dirichlet case.
Using (2.2), (2.3), and (3.1), for each fixed
$x\ge 0$ one can prove that $f(k,x)$ and $f'(k,x)$
in the scalar case satisfy
$$f(-k^\ast,x)^\ast=f(k,x),\quad
f'(-k^\ast,x)^\ast=f'(k,x),\qquad k\in\bCpb.\tag 8.6$$
Informally speaking, $f(k,x)$ and $f'(k,x)$ each contain
$k$ as $ik,$ and hence we have (8.6).
Using (8.6) in (8.5) we see that the Jost matrix in the scalar case
is given by
$$J(k)=f(k,0) \,(\cos\theta)+f'(k,0)
\,(\sin\theta),\qquad k\in\bR,\tag 8.7$$
which is equivalent to
$$J(k)=\cases -f(k,0),\qquad \text{Dirichlet case},\\
\stretch
(\sin\theta)\left[ f'(k,0)+(\cot\theta)\,f(k,0)\right],\qquad \text{non-Dirichlet case}.\endcases\tag 8.8$$
In the literature in the scalar case the Jost matrix is usually
called the Jost function and is defined [7,19,20] as
$$J(k)=\cases f(k,0),\qquad \text{Dirichlet case},\\
\stretch
-i\left[ f'(k,0)+(\cot\theta)\,f(k,0)\right],
\qquad \text{non-Dirichlet case},\endcases\tag 8.9$$
The primary motivation behind the definition in (8.9) is
to define the Jost function $J(k)$ in the scalar case in such a way that
as $k\to\infty$ in $\bCpb$ we have
$J(k)=1+O(1/k)$ in the Dirichlet case and
$J(k)=k+O(1)$ in the non-Dirichlet case.
We remark that (8.8) and (8.9) do not agree, and
we further elaborate on this disagreement.
We know from (b) in Section~3 that the right-hand side of (8.7)
can be multiplied by any nonzero constant because the boundary
matrices $A$ and $B$ appearing in (3.2) can be postmultiplied
by any invertible matrix $T$ without affecting (2.5)-(2.7).
Comparing (8.8) and (8.9) we see that
it is impossible to modify the right-hand side of (8.8)
through a multiplication by a nonzero
scalar so that the right-hand sides of (8.8) and (8.9) agree.
In other words, we cannot use the same multiplicative constant
both in the Dirichlet case and in the non-Dirichlet case so that
(8.8) and (8.9) can agree.

Using (8.8) in
(3.3) we obtain the scattering matrix in the scalar case as
$$S(k)=\cases -\ds\frac{f(-k,0)}{f(k,0)},\qquad \text{Dirichlet case},\\
\stretch
-\ds\frac{f'(-k,0)+(\cot\theta)\,f(-k,0)}
{f'(k,0)+(\cot\theta)\,f(k,0)},\qquad \text{non-Dirichlet case}.\endcases\tag 8.10$$
On the other hand, the scattering matrix in the scalar case is defined in the literature
[7,19,20]
as
$$S(k)=\cases \ds\frac{f(-k,0)}{f(k,0)},\qquad \text{Dirichlet case},\\
\stretch
-\ds\frac{f'(-k,0)+(\cot\theta)\,f(-k,0)}
{f'(k,0)+(\cot\theta)\,f(k,0)},\qquad \text{non-Dirichlet case}.\endcases\tag 8.11$$
Thus, the first lines of (8.10) and (8.11) differ by a minus sign
and their second lines are identical.
Note that (8.9) and (8.11) indicate that the scattering matrix
in the literature [7,19,20] in the scalar case is related to the Jost matrix as
$$S(k)=\cases J(-k)\,J(k)^{-1},\qquad \text{Dirichlet case},\\
\stretch
-J(-k)\,J(k)^{-1},\qquad \text{non-Dirichlet case},\endcases\tag 8.12$$
The definition (8.12) of the scattering matrix in the scalar case
in the literature is motivated by the fact that (8.12)
ensures that $S_\infty$ defined in (4.1) is equal to
$1,$ regardless of the Dirichlet case or the non-Dirichlet case.
Comparing (8.12) with
(3.3) we see that (3.3) and the first line of (8.12)
differ by a minus sign and that (3.3) and the second line
of (8.12) agree with each other.

In the previous literature [1,21], the scattering matrix in the Dirichlet case
is defined as in the first line of (8.12) even in the nonscalar case, i.e.
when $n\ge 2.$
Again, this ensures that $S_\infty=I,$ where we recall that
$I$ is the $n\times n$ identity matrix.
However, defining the scattering matrix in
the Dirichlet case as in (8.12) and
not as (3.3) makes it impossible to have a unique
solution to the inverse scattering problem unless the boundary condition
is already known as a part of the scattering data. If the physical problem
arises mainly from quantum mechanics and hence the boundary condition
is the purely Dirichlet condition, which corresponds to
having $A=0$ in (2.5), this does not present a problem.
On the other hand, if the determination of the selfadjoint
boundary condition is a part of the solution to the inverse problem,
then the definition of the scattering matrix $S(k)$
given in (8.12) is problematic and that is one of the reasons
why we use the
definition of $S(k)$ given in (3.3) regardless of the
boundary condition.
Note that we define the scattering matrix as in (3.3) so that the associated Schr\"odinger operator for the unperturbed problem has the Neumann boundary condition. This definition is motivated by the theory of quantum graphs, where the Neumann boundary condition
is usually used for the unperturbed
problem. We
refer the reader to [15,17,18,22] for
further details.

In the following three examples, we illustrate the drawback of using
(8.12) and not (3.3) as the definition of the scattering matrix.


\noindent {\bf Example 8.1}
Let us use (8.12) as the definition
of the scattering matrix, instead of using (3.3). Let us assume that
we are in the
scalar case. Let us consider the input data set
$\bold D$ given in (2.1) and the scattering data set $\bold S$ given in
(3.12).
The input data set
$\bold D_1$ corresponding to the trivial potential
$V_1(x)\equiv 0$ and the Dirichlet boundary condition with $\theta_1=\pi$
yields the Jost solution
$f_1(k,x)=e^{ikx},$ and hence the corresponding Jost matrix is evaluated
by using the first line of (8.9) as
$J_1(k)=f_1(k,0)=1.$ There are no bound states because
$J_1(k)$ does not vanish on the positive imaginary axis
in the complex $k$-plane. Thus, using the first line of
(8.11), we evaluate the scattering matrix as $S_1(k)\equiv 1,$ and
hence the corresponding scattering data set $\bold S_1$
consists of $S_1(k)\equiv 1$ without any bound states.
On the other hand, the input data set
$\bold D_2$ corresponding to the trivial potential
$V_2(x)\equiv 0$ and the Neumann boundary condition with $\theta_2=\pi/2$
corresponds to the Jost solution
$f_2(k,x)=e^{ikx}$ and hence, by using the second line of
 (8.9), the corresponding Jost matrix is evaluated
as
$J_2(k)=-i f'_2(k,0)=k.$ There are no bound states because
$J_2(k)$ does not vanish on the positive imaginary axis
in the complex $k$-plane. Thus, using the second line of (8.11)
or equivalently using the second line of (8.12),
we evaluate the scattering matrix as $S_2(k)\equiv 1,$ and
hence the corresponding scattering data set $\bold S_2$
consists of $S_2(k)\equiv 1$ without any bound states.
Thus, we have shown that $\bold S_1=\bold S_2$ even though
$\bold D_1\ne \bold D_2.$ This nonuniqueness would not
occur if we used (3.3) as the definition
of the scattering matrix $S(k).$ We would then get
$S_1(k)\equiv -1$ and $S_2(k)\equiv 1,$ and hence
$\bold S_1\ne \bold S_2.$


Next, we further illustrate the nonuniqueness encountered in
Example~8.1 with a nontrivial example.


\noindent {\bf Example 8.2}
Let us use (8.12) as the definition
of the scattering matrix, instead of using (3.3). Let us assume that
we are in the
scalar case. Let us choose a nontrivial potential $V_1(x)$ so that
it is real valued and satisfies (2.4).
Let us also view $V_1(x)$ as a full-line potential
with
support on $x\in\bR^+.$
We refer the reader to
any reference on the scattering theory for the full-line Schr\"odinger equation such as
[2,11-13,19,20] for the description of the corresponding
scattering coefficients.
As a full-line potential, let us also assume that
$V_1(x)$ has no bound states and corresponds to
the full-line exceptional case.
The no bound-state assumption on the full line is the same as
assuming that
the transmission coefficient
has no poles on the positive imaginary axis
in the complex $k$-plane,
and the exceptional case on the full line is equivalent to
the assumption that the transmission coefficient
does not vanish at $k=0.$
Corresponding to $V_1(x)$ as a full-line potential
we have the full-line scattering data
$\{T_1(k),R_1(k),L_1(k)\},$ where
$T_1(k)$ is the transmission coefficient,
$R_1(k)$ is the reflection coefficient from the right,
and $L_1(k)$ is the reflection coefficient from the left.
It is known [2,11-13,19,20] that the full-line scattering data
$\{T_2(k),R_2(k),L_2(k)\},$ where we have
$$T_2(k)=T_1(k),\quad R_2(k)=-R_1(k),\quad L_2(k)=-L_1(k),\tag 8.13$$
corresponds to a nontrivial full-line potential $V_2(x)$
so that
$V_2(x)$ is real valued, vanishes when $x<0,$
has no bound states on the full line, and
corresponds to the full-line exceptional case.
Furthermore, $V_2(x)$ satisfies (2.4).
Viewing $V_1(x)$ and $V_2(x)$ as half-line potentials,
let us now evaluate
the corresponding half-line scattering data sets $\bold S_1$
and $\bold S_2$ associated with
the full-line scattering data sets
$\{T_1(k),R_1(k),L_1(k)\}$ and $\{T_2(k),R_2(k),L_2(k)\},$
respectively.
Since $V_1(x)$ and $V_2(x)$ both vanish when $x<0,$
the corresponding respective Jost solutions $f_1(k,x)$ and
$f_2(k,x)$ yield
$$f_1(k,0)=\ds\frac{1+L_1(k)}{T_1(k)},\quad
f_2(k,0)=\ds\frac{1+L_2(k)}{T_2(k)},\tag 8.14$$
$$f'_1(k,0)=ik\,\ds\frac{1-L_1(k)}{T_1(k)},\quad
f'_2(k,0)=ik\,\ds\frac{1-L_2(k)}{T_2(k)}.\tag 8.15$$
Let us now view $V_1(x)$ as a half-line potential,
associate it with the Dirichlet boundary condition
$\theta_1=\pi,$ and use $\bold D_1$ to denote
the resulting input data set.
 Similarly, let us view $V_2(x)$ as a half-line potential,
associate it with the Neumann
boundary condition $\theta_2=\pi/2,$ and use $\bold D_2$ to denote
the resulting input data set. Clearly, we have $\bold D_1\ne \bold D_2$
because $\theta_1\ne \theta_2.$
Using (8.14) in the
first lines of (8.9) and (8.11) we obtain the Jost matrix
$J_1(k)$ and the scattering matrix $S_1(k)$ as
$$J_1(k)=f_1(k,0)=\ds\frac{1+L_1(k)}{T_1(k)},
\quad S_1(k)=\ds\frac{f_1(-k,0)}{f_1(k,0)}=\ds\frac{T_1(k)}{T_1(-k)}\,
\ds\frac{1+L_1(-k)}{1+L_1(k)}.\tag 8.16$$
On the other hand, using (8.15) and the second lines of (8.9) and (8.11)
with $\theta=\pi/2,$ we obtain the Jost matrix
$J_2(k)$ and the scattering matrix $S_2(k)$ as
$$J_2(k)=-i\,f'_2(k,0)=k\,\ds\frac{1-L_2(k)}{T_2(k)},
\quad S_2(k)=-\ds\frac{f'_2(-k,0)}{f'_2(k,0)}=\ds\frac{T_2(k)}{T_2(-k)}\,
\ds\frac{1-L_2(-k)}{1-L_2(k)}.\tag 8.17$$
Using (8.13) in (8.16) and (8.17) we see that
$S_1(k)\equiv S_2(k),$ and hence the
corresponding scattering data sets $\bold S_1$ and $\bold S_2$
coincide. Thus, we get $\bold D_1\ne \bold D_2$ and
$\bold S_1=\bold S_2.$ This nonuniqueness can be fixed by
using (3.3) and not (8.12) as the definition of
the scattering matrix $S(k).$


The nonuniqueness problem encountered in the previous two examples can also
occur in the nonscalar case, as shown in the following example.
This new example is the generalization of Example~8.2 from the
scalar case to the $n\times n$ matrix case for any positive integer
$n.$ For the relevant scattering theory for the matrix Schr\"odinger equation
on the full line, we refer the reader to [3].


\noindent {\bf Example 8.3}
In this example we assume that $n$ is any positive
integer and not necessarily restricted to $n=1.$ Let us use (8.12) as the definition
of the scattering matrix, instead of using (3.3). Let us again
use (2.1) to describe an input data set
$\bold D$ and use (3.12)
to describe a scattering data set $\bold S$ on the half line.
Consider the class of $n\times n$ matrix-valued
potentials $V(x)$ on the full line satisfying
$$V(x)=V(x)^\dagger, \qquad x\in\bR,\tag 8.18,$$
$$\int_{-\infty}^\infty dx\, (1+|x|)\,|V(x)|<+\infty,\tag 8.19$$
where we recall that the dagger denotes the matrix adjoint and
$|V(x)|$ denotes the matrix operator norm. The reader is referred to
[3] for the matrix-valued scattering coefficients
 for the full-line
matrix Schr\"odinger equation with such potentials. Associated
with $V(x)$ we have the full-line scattering coefficients
$T_{\text{l}}(k),$ $R(k),$ and $L(k),$ each of which
is an $n\times n$ matrix. These matrix-valued scattering
coefficients are the matrix
generalizations of the scalar scattering coefficients $T(k),$
$R(k),$ and $L(k)$
considered in Example~8.2.
Let us further restrict the full-line potentials $V(x)$ so that they vanish when
$x<0,$ they do not possess any bound states on the full line, and
they correspond to the purely exceptional case. We refer the reader to [3] for
the details on the bound states and the purely exceptional case on the full line.
The absence of bound states on the full line is equivalent to having the determinant
of the matrix inverse of $T_{\text{l}}(k)$ not vanishing on the
positive imaginary axis in the complex $k$-plane.
The purely exceptional case on the full line is equivalent to
having
the limit of $k\,T_{\text{l}}(k)^{-1}$ as $k\to 0$
equal to the $n\times n$ zero matrix. For such potentials
$T_{\text{l}}(0)^{-1}$ is well defined, and we have $\det[I\pm L(0)]\ne 0,$
where we recall that $I$ denotes the $n\times n$ identity matrix.
Since we only consider the full-line potentials $V(x)$ vanishing when
$x<0,$ we can view their restrictions
on $x\in\bR^+$ as half-line potentials $V(x).$ From (8.18) and (8.19)
we see that their restrictions on $x\in\bR^+$ belong to the Faddeev class.
When $x\in\bR^+,$ the full-line Jost solution from the left $f_{\text{l}}(k,x)$
coincides [3] with the half-line Jost solution $f(k,x)$ appearing
in (3.1). Furthermore, we have [3]
$$f_{\text{l}}(k,0)=[I+L(k)]\,T_{\text{l}}(k)^{-1},\quad
f'_{\text{l}}(k,0)=ik\, [I-L(k)]\,T_{\text{l}}(k)^{-1},\qquad k\in\bR.\tag 8.20$$
Let $V_1(x)$ be a specific full-line matrix potential
satisfying (8.18) and (8.19) such that
it vanishes when $x<0,$ does not contain any bound states on the full line, and corresponds to
the purely exceptional case on the full line.
Let $\{T_{\text{l}}(k),R(k),L(k)\}$ be the corresponding
full-line scattering data. Let $V_2(x)$ be the full-line matrix potential
corresponding to the full-line scattering data
$\{T_{\text{l}}(k),-R(k),-L(k)\},$
where the signs of the matrix-valued reflection coefficients are changed.
The matrix potential $V_2(x)$ also vanishes for $x<0,$ satisfies
(8.18) and (8.19), does not possess any bound states
on the full line, and corresponds to a purely exceptional case
on the full line. The restrictions of
$V_1(x)$ and $V_2(x)$ on $x\in\bR^+$ can be viewed as
half-line potentials. Let $A_1,$ $B_1,$ $A_2,$ $B_2$ be four
$n\times n$ constant matrices in such a way that
$A_1=0,$ $B_2=0,$ $A_2$ is an arbitrary invertible matrix, and $B_1$ is an
arbitrary invertible matrix.
Let $\bold D_1:=\{V_1,A_1,B_1\}$ and
$\bold D_2:=\{V_2,A_2,B_2\}$ be the half-line input data sets
as in (2.1), with the understanding that
the domains of $V_1(x)$ and $V_2(x)$ are restricted to $x\in\bR^+.$
Let $f_1(k,x)$ and $f_2(k,x)$ be the half-line Jost solutions
corresponding to $\bold D_1$ and $\bold D_2,$ respectively.
 From (8.20) we see that
$$f_1(k,0)=[I+L(k)]\,T_{\text{l}}(k)^{-1},\quad
f'_1(k,0)=ik\, [I-L(k)]\,T_{\text{l}}(k)^{-1},\qquad k\in\bR,\tag 8.21$$
$$f_2(k,0)=[I-L(k)]\,T_{\text{l}}(k)^{-1},\quad
f'_2(k,0)=ik\, [I+L(k)]\,T_{\text{l}}(k)^{-1},\qquad k\in\bR.\tag 8.22$$
Using (8.21) and (8.22), because of the purely
exceptional case on the full line [3],
it follows that
neither of the determinants of $f_1(k,0)$ and $f'_2(k,0)$ vanish.
Let $J_1(k),$ $S_1(k),$ $\bold S_1$
be the respective Jost matrix, scattering matrix, and scattering data set
corresponding to $\bold D_1.$
Similarly, let
$J_2(k),$ $S_2(k),$ $\bold S_2$
be the respective Jost matrix, scattering matrix, and scattering data set
corresponding to $\bold D_2.$
Using (8.21) and (8.22) in (3.2) we obtain
$$J_1(k)=f_1(-k,0)^\dagger\, B_1=[T_{\text{l}}(-k)^\dagger]^{-1}\, [I+L(-k)^\dagger ]\,B_1,
\tag 8.23$$
$$J_2(k)=-f'_2(-k,0)^\dagger A_2=-ik\, [T_{\text{l}}(-k)^\dagger]^{-1}\, [I+L(-k)^\dagger ]\,A_2.\tag 8.24$$
Using (8.23) in the first line
of
(8.12) we obtain
$$S_1(k)=J_1(-k)\,J_1(k)^{-1}=f_1(k,0)^\dagger
 \left[f_1(-k,0)^\dagger\right]^{-1},$$
 which yields
$$S_1(k)=[T_{\text{l}}(k)^\dagger]^{-1}\, [I+L(k)^\dagger ]\,[I+L(-k)^\dagger ]^{-1}\,T_{\text{l}}(-k)^\dagger.\tag 8.25$$
Using (8.24) in the second line
of
we obtain
$$S_2(k)=-J_2(-k)\,J_2(k)^{-1}=-f'_2(k,0)^\dagger\,
 \left[f'_2(-k,0)^\dagger\right]^{-1},$$
yielding
$$S_2(k)=[T_{\text{l}}(k)^\dagger]^{-1}\, [I+L(k)^\dagger ]\,[I+L(-k)^\dagger ]^{-1}\,T_{\text{l}}(-k)^\dagger.\tag 8.26$$
There are no half-line bound states associated with either
of the scattering data sets corresponding to
$S_1(k)$ and $S_2(k).$ Hence we have
$\bold S_1=\{S_1\}$ and $\bold S_2=\{S_2\}.$
 From (8.25) and (8.26) it follows that
 $S_1(k)\equiv S_2(k)$ and hence we have
 $\bold S_1=\bold S_2$ even though
 $\bold D_1\ne \bold D_2.$
This nonuniqueness would not
occur if we used (3.3) as the definition
of the scattering matrix $S(k).$ We would then get
$S_1(k)\equiv -S_2(k),$ and hence
$\bold S_1\ne \bold S_2.$


\vskip 20 pt

\noindent {\bf Acknowledgments.}
The research leading to this
article was supported in part by CONACYT under project CB2015,
254062, Project PAPIIT-DGAPA-UNAM IN103918, and by
Coordinaci\'on de la Investigaci\'on Cient\'{\i}fica, UNAM.

\vskip 20 pt

\noindent {\bf{References}}

\vskip 3 pt

\item{[1]} Z. S. Agranovich and V. A. Marchenko, {\it The inverse problem of
scattering theory,} Gordon and Breach, New York, 1963.

\item{[2]} T. Aktosun and M. Klaus,
{\it Chapter 2.2.4: Inverse theory: problem on the line,}
In: E. R. Pike and P. C. Sabatier (eds.),
{\it Scattering,}
Academic Press, London, 2001, pp. 770--785.

\item{[3]} T. Aktosun, M. Klaus, and C. van der Mee,
it{Small-energy asymptotics for the
Schr\"odinger equation on the line,}
J. Math. Phys. {\bf 17}, 619--632 (2001).

\item{[4]} T. Aktosun, M. Klaus, and R. Weder,
{\it Small-energy analysis for the self-adjoint matrix
Schr\"odinger operator on the half line,}
J. Math. Phys. {\bf 52}, 102101 (2011);
arXiv:1105.1794 [math-ph] (2011).

\item{[5]} T. Aktosun, M. Klaus, and R. Weder,
{\it Small-energy analysis for the self-adjoint matrix
Schr\"odinger operator on the half line. II,}
J. Math. Phys. {\bf 55}, 032103 (2014);
arXiv:1310.4809 [math-ph] (2013).

\item{[6]} T. Aktosun, P. Sacks, and M. Unlu,
{\it Inverse problems for selfadjoint Schr\"odinger operators on the half line with compactly supported potentials,}
J. Math. Phys. {\bf 56}, 022106 (2015);
arXiv:1409.5819 [math.SP] (2014).

\item{[7]} T. Aktosun and R. Weder,
{\it Inverse spectral-scattering problem with two sets
of discrete spectra for the radial Schr\"odinger equation,}
Inverse Problems {\bf 22}, 89--114 (2006);
arXiv:math-ph/0402019 (2004).

\item{[8]} T. Aktosun and R. Weder,
{\it High-energy analysis and Levinson's theorem for the self-adjoint matrix Schr\"odinger operator on the half line,}
J. Math. Phys. {\bf 54}, 112108 (2013);
arXiv:1206.2986 [math-ph] (2012).

\item{[9]} T. Aktosun and R. Weder,
{\it Inverse scattering for the matrix Schr\"odinger equation,} preprint, 2018;
arXiv:1708.03837 [math-ph] (2017).

\item{[10]} T. Aktosun and R. Weder,
{\it Direct and inverse scattering for the matrix Schr\"odinger equation,}
the monograph to be published by Springer-Verlag.

\item{[11]} K. Chadan and P. C. Sabatier, {\it Inverse problems in quantum
scattering theory,} 2nd ed., Springer, New York, 1989.

\item{[12]} P. Deift and E. Trubowitz, {\it Inverse scattering
    on the line,} Commun. Pure Appl. Math. {\bf 32}, 121--251
    (1979).

\item{[13]} L. D. Faddeev, {\it Properties of the $S$-matrix of
    the one-dimensional Schr\"odinger equation,} Amer. Math.
    Soc. Transl. {\bf 65} (ser. 2), 139--166 (1967).

\item{[14]} M. S. Harmer, {\it Inverse scattering for the matrix Schr\"odinger
operator and Schr\"odinger operator on
graphs with general self-adjoint boundary conditions,}
ANZIAM J. {\bf 44}, 161--168 (2002).

\item{[15]} M. S. Harmer, {\it The matrix Schr\"odinger operator
and Schr\"odinger operator on graphs,} Ph.D. thesis, University of
Auckland, New Zealand, 2004.

\item{[16]} M. Harmer, {\it Inverse scattering on matrices with boundary conditions,}
J. Phys. A {\bf 38}, 4875--4885 (2005).

\item{[17]} V. Kostrykin and R. Schrader,
{\it Kirchhoff's rule for quantum wires,} J. Phys. A {\bf 32}, 595--630
(1999).

\item{[18]} V. Kostrykin and R. Schrader,
{\it Kirchhoff's rule for quantum wires. II: The inverse problem with possible applications to quantum computers,} Fortschr. Phys. {\bf 48}, 703--716
(2000).

\item{[19]} B. M. Levitan, {\it Inverse Sturm-Liouville problems,}
VNU Science Press, Utrecht, 1987.

\item{[20]}  V.\; A.\; Marchenko, {\it Sturm-Liouville operators and
applications,} revised ed., Amer. Math. Soc. Chelsea Publ., Providence, R.I., 2011.

\item{[21]} R. G. Newton and R. Jost,
{\it The construction of potentials from the $S$-matrix
for systems of differential equations,} Nuovo Cim. {\bf 1}, 590--622
(1955).

\item{[22]} R. Weder,
{\it Scattering theory for the matrix Schr\"odinger operator on the half line with general boundary conditions,}
J. Math. Phys. {\bf 56}, 092103 (2015);
arXiv:1505.0879 [math-ph] (2015).

\item{[23]} R. Weder,
{\it Trace formulas for the matrix Schr\"odinger operator on the half-line with general boundary conditions,}
J. Math. Phys. {\bf 57}, 112101 (2016);
arXiv:1603.09432 [math-ph] (2016).

\item{[24]} R. Weder,
{\it The number of eigenvalues of the matrix Schr\"odinger operator on the half line with general boundary conditions,} J. Math. Phys. {\bf 58}, 102107 (2017);
arXiv:1705.03157 [math-ph] (2017).

\end